\documentclass[useAMS,usenatbib]{mn2e}
\usepackage{graphicx}
\usepackage{epstopdf}
\usepackage[figuresright]{rotating}
\usepackage{subfigure,color}
\usepackage{longtable}
\usepackage{xspace}

\newcommand{\ion}[2]{{\textrm{#1}}\,{\textrm{\sc #2}}}

\title[Effective temperature of  ionizing stars of extragalactic \ion{H}{ii} regions]
{Effective temperature of  ionizing stars of extragalactic \ion{H}{ii} regions}
\author[Dors et al.]
{O.~L. Dors$^{1}$\thanks{E-mail:\,olidors@univap.br}, G.~F. H\"agele$^{2,3}$, M.~V. Cardaci$^{2,3}$, A.~C. Krabbe$^{1}$\\
$^1$ Universidade do Vale do Para\'iba, Av. Shishima Hifumi, 2911, Cep12244-000, S\~ao Jos\'e dos Campos, SP, Brazil\\
$^2$ Instituto de Astrof\'isica de La Plata (CONICET-UNLP), Argentina. \\
$^3$ Facultad de Ciencias Astron\'omicas y Geof\'{\i}sicas, Universidad Nacional de La Plata, Paseo del Bosque s/n, 1900 La Plata, Argentina.\\
}

\begin{document}

\date{Accepted- 2011 April 28. Received -2011 February 18.}

\pagerange{\pageref{firstpage}--\pageref{lastpage}} \pubyear{2011}

\maketitle

\label{firstpage}

\begin{abstract}

The effective temperature ($T_{\rm eff}$) of the radiation field of the ionizing star(s)  of a large sample of extragalactic \ion{H}{ii} regions was
estimated using the  $R$=log([\ion{O}{ii}]($\lambda$$\lambda$3726+29)/[\ion{O}{iii}]$\lambda$5007) index.
We used a grid of photoionization models to calibrate the $T_{\rm eff}$-$R$ relation finding that it has a strong dependence with the ionizing parameter while it shows a weak  direct dependence with the metallicity (variations in $Z$ imply variations in $U$) of both the stellar atmosphere of the ionizing star and the gas phase of the \ion{H}{ii} region.
 Since the $R$ index varies slightly with the $T_{\rm eff}$ for values larger than 40\,kK, the $R$ index can be used to derive the $T_{\rm eff}$ in the $30-40$ kK range.
A large fraction of the ionization parameter variation is due to differences in the temperature of the ionizing stars and then 
the use of the (relatively) low $T_{\rm eff}$ dependent 
$S2$=[\ion{S}{ii}]($\lambda\lambda$6717+31)/H$\alpha$ emission-line ratio
to derive the ionization parameter is preferable over others in the literature.
 We propose linear metallicity dependent relationships between $S2$ and $U$.
$T_{\rm eff}$ and metallicity estimations for a sample of 865 \ion{H}{ii} regions, whose emission-line intensities 
were compiled from the literature, do not show any $T_{\rm eff}$-$Z$ correlation.  On the other hand it seems to be hints of the presence of an anti-correlation between $T_{\rm eff}$-$U$.
We found that the majority of the studied \ion{H}{ii} regions ($\sim 87$\%) present $T_{\rm eff}$  values in the range between $37$ and $40$\,kK, 
with an average value of  $38.5 (\pm 1)$ kK. 
We also studied the variation of $T_{\rm eff}$ as a function of the galactocentric distance for 14 spiral galaxies. Our results are in agreement with the idea of the existence of positive $T_{\rm eff}$ gradients along the disk of spiral galaxies.

 \end{abstract}



\section{Introduction}
The effective temperature of massive stars is an important parameter to understand the evolution of these objects and their 
influence on the interstellar medium as well as on the galaxy in which they reside.

In general, the effective temperature of ionizing stars of an \ion{H}{ii} region (hereafter $T_{\rm eff}$) located in the Milk Way
and in the Magellanic Clouds can be estimated  through their spectral classifications (see e.g.\ \citealt{evans15, lamb15, walborn14, 
morrell14, sota14, sota11, massey09, massey05, conti07}).
However, for  more distant stars, the $T_{\rm eff}$ can be only estimated 
indirectly, i.e.\ through  the analysis of  the emission-lines emitted by the nebulae ionized by these stars. The original idea was 
proposed by \citet{zanstra31} and consist in assuming that, if the nebula around of a star is optically thick to the Lyman continuum, 
it absorbs all  the ionizing photons emitted by the star. Thus, the number of ionizations per unit  of time in the nebula or the flux of a given emission-line
is directly proportional to the number of ionizing photons emitted by the star as well as  dependent  on its  effective temperature (\citealt{osterbrock89}). 
Along decades,  this method have been improved
by several authors, mainly  in the sense of defining what nebular lines
can be used to constrain $T_{\rm eff}$   in Planetary Nebulae and  
in  \ion{H}{ii} regions  (e.g.\ \citealt{ambarzumian32, stoy33, gurzadyan55, kaler76, chopinet76, kaler78, koppen78, iijima81,
mathis85, stasinska86, vilchez88, bresolin99, kennicutt00, oey00, dors03, morisset04, enrique09a, zanstrow13}).

$T_{\rm eff}$  determinations for  the majority of
extragalactic  \ion{H}{ii} regions are obtained comparing 
observed emission-line  fluxes with those predicted
by nebular photoionization models assuming as ionizing source a star with a given temperature 
(e.g.\ \citealt{zanstrow13, pellegrini11, morisset04, dors03, kennicutt00, vilchez88}). 
 However,    nebular emission-lines depend primarily on three parameters:
$T_{\rm eff}$, metallicity and ionization
parameter  \citep{oey00}. Thus, it is necessary produce estimations of two of these
parameters to derive the  third. \citet{kennicutt00},
who used optical data of \ion{H}{ii} regions located  in the
Milk Way and  the Magellanic Clouds,
compared  $T_{\rm eff}$ estimations based on  emission-lines predicted by photoionization models with values obtained from stellar spectral classifications.  These authors 
observed a strong degeneracy between forbidden-line sequences produced by changes in  
$T_{\rm eff}$ and metallicity of the gas phase of the nebulae, which shows the  difficulty of using emission-line ratios as $T_{\rm eff}$ indicators.

Likewise, \citet{morisset04},
  who compared  diagnostic diagrams containing observational infrared emission-lines of Galactic \ion{H}{ii} regions and  predictions  from  photoionization models, 
showed that if the metallicity and the ionization parameter
of the gas phase of the nebulae are not taking into account, erroneous $T_{\rm eff}$ values can be obtained.
 Moreover, \citet{pellegrini11} showed that $T_{\rm eff}$ can be determined
through sets  of diagnostic  diagrams containing   photoionization model predictions of emission lines  dependent on the 
radiation  flux emitted by the  ionizing source and on the ionizing parameter, but weakly
dependent  on the metallicity and the electron density of the gas. This methodology permits
  to break the degeneracy in $T_{\rm eff}$ estimations.  Most recently, \citet{zanstrow13} compared
long slit observations of  a sample of \ion{H}{ii} regions located in  the Large Magellanic Cloud with predictions  
of  photoionization models  in order to estimate the $T_{\rm eff}$  of these objects.
 These authors pointed out the need in estimating the metallicity 
 and the ionizing parameter before using the models to
 calculate the  $T_{\rm eff}$.

To eliminate the degeneracy in the $T_{\rm eff}$ estimations
it is required  to calculate the metallicity ($Z$) of the gas phase  
(generally traced by the relative abundances between oxygen and hydrogen, O/H)
and the ionization parameter of the \ion{H}{ii} regions considered.  Concerning the first parameter,
accurate  metallicities  of   \ion{H}{ii} regions can only be derived by measuring auroral
emission-lines (e.g.\ [\ion{O}{iii}]$\lambda$4363)
which are  very weak or unobservable  in  \ion{H}{ii} regions with high
metallicity and/or low excitation  \citep{bresolin05,diaz07}. Therefore, 
for most of the cases, the method  based on
calibration between strong emission-lines and oxygen abundances  proposed by  \citet{pagel79}
are used to estimate $Z$ (see also 
\citealt{leonid12, pena12, dors05, kenniccut03, kewley02, leonid01}). 
Concerning   the ionizing parameter $U$, \cite{diaz91} and \cite{dors11}
 derived calibrations between emission-line ratios easily  measurable
and $U$, which can be used to eliminate the degeneracy in the
 $T_{\rm eff}$ estimations.

Nowadays, despite the large amount of  spectroscopic data of \ion{H}{ii} regions available in the literature, such as the data produced by the CALIFA survey \citep{sanchez12}, $T_{\rm eff}$ 
 has  been estimated for few extragalactic objects. In fact, \citet{kennicutt00} estimated the  $T_{\rm eff}$ for 39 \ion{H}{ii} regions, being only 10 objects located in the Magellanic Clouds. 
 \citet{dors03}, using the line ratio  $R=$log([\ion{O}{ii}]($\lambda$$\lambda$3726+29)/[\ion{O}{iii}]$\lambda$5007) and  the spectroscopic data of \cite{kennicutt96},
derived  $T_{\rm eff}$ values for exciting stars located in the disk of the spiral galaxy M\,101 (see also \citealt{evans86, vilchez88, henry95, zanstrow13}).
 
In this paper we  also used the   $R$=log([\ion{O}{ii}]($\lambda$$\lambda$3726+29)/[\ion{O}{iii}]$\lambda$5007) index to estimate the 
$T_{\rm eff}$ for a large sample of extragalactic \ion{H}{ii} regions. Our study is motivated by
the following goals: \\
(i)  To produce a calibration between $R$ and $T_{\rm eff}$ taking into account 
the effects of the ionizing parameter and the metallicity of both nebular gas and stellar atmosphere of the ionizing star on this relation. \\
(ii)  To estimate $T_{\rm eff}$ values  for a large sample
of extragalactic \ion{H}{ii} regions and investigate the dependence of $T_{\rm eff}$ with the metallicity and  with the ionization parameter.\\
(iii) To  investigate the variation of  $T_{\rm eff}$  
with the galactocentric distance in  spiral galaxies
in order to verify if gradients of this parameter are an universal
property of these objects.
 
The photoionization models used to obtain 
an $R$-$T_{\rm eff}$ calibration are described in Sect.~\ref{mod}.
   The data sample used to derive $T_{\rm eff}$ values are presented in Sect.~\ref{obs}.
 The  methodology employed  and the  sources of uncertainties
 are described  in Sects.~\ref{met} and \ref{err}, respectively.
  In Sect.~\ref{res} and \ref{disc} the results and  discussion of the outcome
 are presented. The 
final conclusions are given in Sect.~\ref{conc}.

 
\section{Photoionization models}
\label{mod}

We employed the {\sc Cloudy} code version 13.00 \citep{ferland13}
to build a grid of photoionization models in order
to derive calibrations among the parameters
$T_{\rm eff}$ and $U$ and  strong nebular  
emission-lines, easily  measurable in observations 
of \ion{H}{ii} regions. In what follows the main parameters
of the models are described.
\begin{itemize}
 \item Metallicity -- The metallicity $Z$ of the gas phase of the hypothetical 
nebulae was linearly scaled  with the solar metallicity $Z_{\odot}$, considering
the solar oxygen abundance $\rm 12+\log(O/H)_{\odot}=8.69$ \citep{alendeprieto}.
 The nitrogen abundance  was taken from  the relation
$\rm \log (N/O) = log (0.034 + 120 \: O/H)$ of \citet{vilacostas93}.
We considered the values  $Z$=1.0, 0.5, 0.2 and 0.03  $Z_{\odot}$.
The presence of internal dust was considered and the grain abundances 
\citep{vanhoof01} were also linearly scaled with  $Z$.
 Depletion of refractory elements onto dust grains was considered
 as in \citet{dors05}.
 \item Electron density -- We considered the electron density 
 as being $100 \: \rm cm^{-3}$. This value is in the range 
 of values derived for extragalactic \ion{H}{ii} regions 
 \citep{sanders16, krabbe14, coppeti00}.
 
 \item  Stellar atmosphere model -- We use the public
stellar atmosphere models WM-basic \citep{pauldrach01}
that are already available in the stellar atmosphere library in the
{\sc Cloudy} code \citep{ferland13}. We considered the WM-basic  models because, among the 
stellar atmosphere models assumed in the   photoionization
model grids built by \citet{zanstrow13}, they  produced  the best agreement
between predicted  and  observed optical emission-line ratios.
The $T_{\rm eff}$ values ranged 
from 30\,000 to 50\,000 K, with a step of 2500 K, where the
 metallicity ($Z$) of the stellar atmosphere was considered
 to be  the same  than the one of the nebular gas. This $T_{\rm eff}$ range   is the same that the one considered by \citet{morisset04}.
 
  The value of the solar metallicity is a  current question of debate (see \citealt{caffau16}) and it is constantly updated in the {\sc Cloudy} code\footnote{see also $http://www.nublado.org/wiki/RevisionHistory$}, which
 can yield an  incorrect match between the abundances
 of  some particular  element  in the gas phase of the nebulae and in 
 the stellar atmosphere.  Unfortunately, this problem can not be  
 resolved because  the assumed abundance values for
 the majority of the elements in stellar atmosphere models are, many times, not declared and only a general value of the metallicity is given. 
  
 \item Ionization parameter -- It is defined as $U={Q_{\rm ion}}/4\pi R^{2}_{\rm in} n  {\rm c}$, where ${Q_{\rm ion}}$  
is the number of hydrogen ionizing photons emitted per second
by the ionizing source, $R_{\rm in}$  is  the distance from the ionization source to the inner surface
of the ionized gas cloud (in cm), $n$ is the  particle
density (in $\rm cm^{-3}$), and $\rm c$ is the speed of light. 
We considered $\log U$ ranges from $-3.5$ to $-1.5$ dex, with a step of $0.5$ dex.
The variation in the value of $U$ simulates  
the excitation differences  of  \ion{H}{ii} regions, mass,
or geometrical conditions in a wide range of possible scenarios
\citep{perez14}.  It is worth mentioning that models with different combination of $Q_{\rm ion}$, $R$ and $n$ 
but that result in the same $U$ are homologous models with
the same predicted emission-line intensities \citep{bresolin99}.

 \end{itemize}
 
In  total, 180 photoionization models were built. The $T_{\rm eff}$ value
estimated  through the $R$ index should be interpreted as the  
temperature of the hottest star of  the ionizing stellar cluster  of  the \ion{H}{ii} region, since this star
drives the emission of the ionizing photons  \citep{zanstrow13}.

 
\section{Data Sample}
\label{obs}

Observational emission-line intensities of a sample of extragalactic
  \ion{H}{ii} regions  were compiled from the literature. 
We considered 
only \ion{H}{ii} regions  for which the intensities, relative to H$\beta$, of the [\ion{O}{ii}]$\lambda\lambda$3726+29,
[\ion{O}{iii}]$\lambda$5007, H$\alpha$, and [\ion{N}{ii}]$\lambda$6584, and [\ion{S}{ii}]$\lambda$$\lambda$6717+31  emission-lines
were measured. All emission line intensities are reddening corrected.
For some few cases in which the H$\alpha$
intensity is not presented, we calculated it 
from  the theoretical ratio  H$\alpha$/H$\beta=2.86$ \citep{hummer97}.
Indeed, when only the sum of [\ion{O}{iii}]($\lambda$4959+$\lambda$5007) 
and/or  [\ion{N}{ii}]($\lambda$6548+$\lambda$6584) 
are listed in the original papers  from which  the data were compiled,
the intensities of [\ion{O}{iii}]$\lambda$5007 and [\ion{N}{ii}]$\lambda$6584 were calculated assuming the theoretical 
relations [\ion{O}{iii}]$\lambda$5007$\approx \: 3.0 \: \times$ [\ion{O}{iii}]$\lambda$4959
and  [\ion{N}{ii}]$\lambda$6584$\approx \: 3.00 \: \times $ [\ion{N}{ii}]$\lambda$6548 \citep{storey00}, respectively.

To exclude objects with  a secondary ionizing source,
we use the criterion proposed by \citet{kewley01} to separate objects 
ionized by massive stars from those containing gas shock and/or
active galactic nuclei (AGN), where all objects with 
\begin{equation}
\label{eq4}
\rm log([O\:III]\lambda5007/H\beta) \: > \: \frac{0.72}{log([S\:II](\lambda\lambda6717+31)/H\alpha)-0.32}+1.30
\end{equation}
were not considered in our sample. 
\begin{figure}
\centering
\includegraphics[angle=-90,width=1\columnwidth]{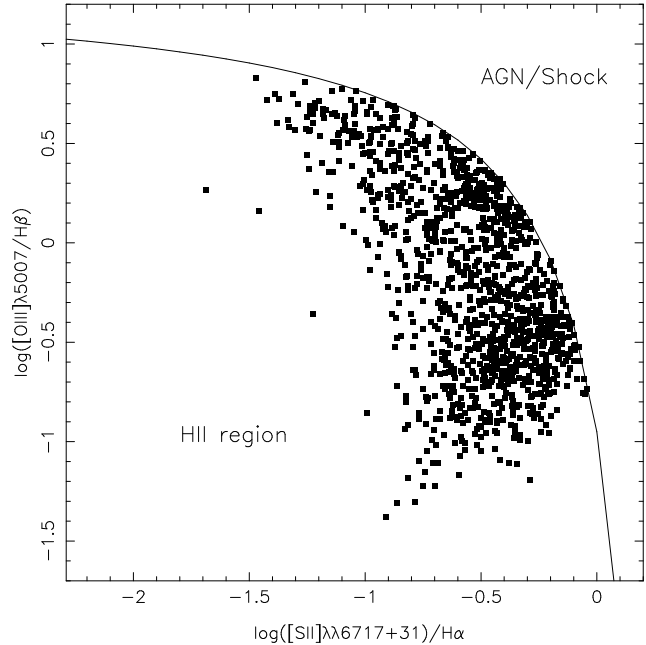}
\caption{log([\ion{O}{iii}]$\lambda$5007/H$\beta$) vs.\ log([\ion{S}{ii}]($\lambda \lambda$6717+31)/H$\alpha$) diagnostic diagram.
The solid line  represents the relation by \citealt{kewley01}. It separates objects ionized by massive stars from those 
containing active nuclei and/or shock-excited gas, as indicated.  Squares represent the observational
data (see Sect.~\ref{obs}).}
\label{f5}
\end{figure} 
We selected  1198 extragalactic \ion{H}{ii} regions located in 44 galaxies with redshift $z<0.5$.\footnote{The redshift values 
were obtained from the NASA/IPAC Extragalactic Database (NED) http://ned.ipac.caltech.edu/} In Table~\ref{tab2} we listed the 
bibliographic references of the sample, the host 
galaxy of the \ion{H}{ii} regions  and the number of objects taken from each work.  
In Fig.~\ref{f5}, a  diagnostic diagram log([\ion{O}{iii}]$\lambda$5007/H$\beta$) versus log([\ion{S}{ii}]$\lambda$$\lambda$6717+31+/H$\beta$)
proposed by \citet{veilleux87}, the observational data and the relation by \citet{kewley01} are shown.

\begin{table*}
\centering
\caption{The sample.}
\label{tab2}
\vspace{0.2cm}
\begin{tabular}{@{}lcc@{}}
\hline
Reference               &        Galaxy                      &     Number of   \ion{H}{ii} regions  \\	 
\hline
\citet{kennicutt00}    &     Magellanic Clouds       &   8   \\
 \citet{vermeij02}     &     Magellanic Clouds       &   6 \\
 \citet{russel90}       &     Magellanic Clouds       &   5 \\
 \citet{zurita12}       &         M\,31                      &   12 \\
\citet{garnett97}      &       NGC\,2403               &    8 \\
\citet{vanzee00}      &        UGCA\,292              &   2\\
\citet{bresolin04}     &     M\,51                          &  10\\
\citet{kenniccut03}  &     M\,101                         &  25 \\
\citet{vilchez88a}    &     M\,33                           & 5 \\
\citet{kwitter81}      &    M\,33                            & 10 \\
\citet{lopezhernadez13} & M\,33                         &  9 \\
\citet{bresolin09a}   &      NGC\,300                   &  16 \\   
\citet{lee04}           &      NGC\,1705                  &  13 \\
\citet{lopezsanchez09} &   Mkn\,1199                 &  5 \\
\citet{lopezsanchez09} & Mkn\,5                        & 2 \\
\citet{lopezsanchez09} &  IRAS\,08208+2816     & 4 \\
\citet{lopezsanchez09} & III\,Zw107                   & 3\\
\citet{lopezsanchez09} &   Tol 1457-262            & 1 \\
\citet{lopezsanchez11} &    IC\,10                     & 7 \\
\citet{lopezsanchez07} &    NGC\,5253             & 4 \\
 \citet{esteban99}       &    Mkn\,8                     & 5 \\
 \citet{berg13}             &    NGC\,628               &  13 \\
\citet{bresolin09b}      &      M\,83                    & 24\\
\cite{bresolin12}         &     NGC\,1512            &  50 \\
\cite{bresolin12}         &     NGC\,3621            &   71 \\
 \citet{vanzee98}         &    NGC\,925              &   24 \\
 \citet{vanzee98}         &   NGC\,1068             & 1 \\
 \citet{vanzee98}         & NGC\, 1232             &  16 \\
 \citet{vanzee98}         & NGC\,1637             & 16 \\
 \citet{vanzee98}        &  NGC\,2805           &  17 \\
\citet{vanzee98}       &  IC\,2458               &  3 \\
\citet{vanzee98}       &   NGC\,2820           & 4 \\
\citet{vanzee98}       &   NGC\,2903            & 9 \\
\citet{vanzee98}      &  NGC\,3184           & 17 \\
\citet{vanzee98}      & NGC\,4395            & 9 \\
\citet{hagele12}      &   Haro \,15             & 2 \\
 \citet{hagele11}      &  SDSS J165712.75+321141.4  &    3 \\
 \citet{diaz07}         & NGC2903              & 4 \\
 \citet{sanchez12}   &   UGC\, 9837         & 64 \\
  \citet{sanchez12}   & NGC\,1058           &  258    \\
 \citet{sanchez12}   &  UGC\,9965          & 56 \\
 \citet{sanchez12}   &   NGC\, 1637        & 148 \\
 \citet{sanchez12}   &  NGC\,3184          & 58 \\
 \citet{sanchez12} & NGC\,3310            & 103 \\
\citet{sanchez12}  & NGC\,4625            & 42 \\
\citet{sanchez12}   & NGC\,5474           & 79 \\
\citet{sanchez12}   & NGC\,628           & 165 \\ 
\hline\noalign{\smallskip}
\end{tabular}
\end{table*}

\section{Methodology}
\label{met}


\subsection{$T_{\rm eff}$ estimation}

To estimate the $T_{\rm eff}$ value  of  a given \ion{H}{ii} region,  we adopted a similar methodology
to the one presented by \citet{dors03}, in which 
 photoionization  model results were considered to derive a relation between $T_{\rm eff}$ and the 
 $R=\log($[\ion{O}{ii}]($\lambda$$\lambda$3726+29)/[\ion{O}{iii}]$\lambda$5007) index.
However, \citet{dors03} did not present an expression to derive $T_{\rm eff}$ 
since they only considered  models with  $Z_{\odot}$ and  with a fixed ionization parameter 
value ($<\log U>=-2.5$). In this work, the $Z$ and $U$ parameters are taking into account in $T_{\rm eff}$ estimations.
   
\begin{figure}
\centering
\includegraphics[angle=-90,width=1\columnwidth]{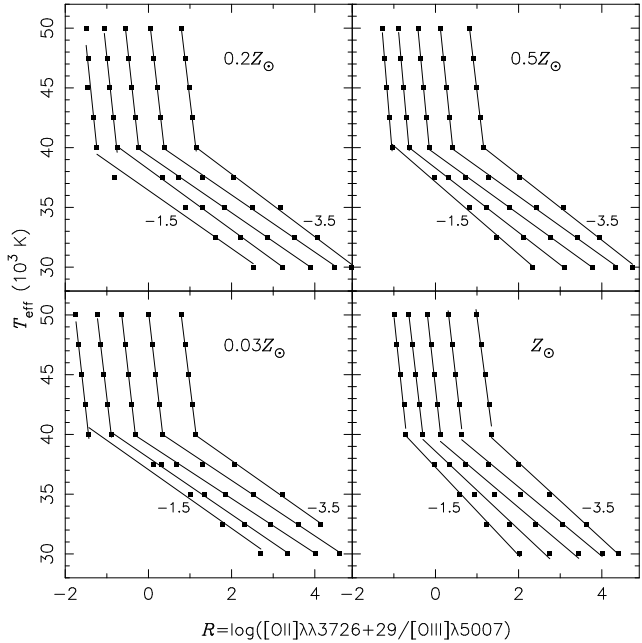}
\caption{$T_{\rm eff}$ vs.\ $R$  assuming different values for the metallicity $Z$ and the ionizing parameters $U$ as indicated. 
Points represent results of the photoionization models (see Sect.~\ref{mod}). Lines represent the linear regression fittings whose 
coefficients are listed in Table~\ref{tab1}.}
\label{f1}
\end{figure}

In Fig.\ \ref{f1},  the relations between $T_{\rm eff}$ and $R$ for  different $Z$ and $U$ values are shown. 
As can be seen, the $T_{\rm eff}$-$R$ relation presents two
behaviours for the ranges of values $T_{\rm eff}$= 30-40 
and 40-50 ($10^{3}$ K). Therefore, we considered different linear regressions for these ranges whose coefficients are listed in Table~\ref{tab1}.
In  Fig.~\ref{f1} we can note the strong dependence of the 
$T_{\rm eff}$-$R$ relation with the ionization parameter. In opposite, the metallicity has a secondary influence on this relation.
 Moreover, we can see that the $R$ index presents little variations for  $T_{\rm eff} \: > \: 40$ kK, result also found 
by \cite{oey00} and \citet{kennicutt00} for  other line-ratios. Hence, for temperatures higher than 40\,kK, small emission-line variations or measurement errors translate into very large uncertainties in the $T_{\rm eff}$ estimations. Thus, despite the $T_{\rm eff}$-$R$ fitting results
are presented in  Table~\ref{tab1}, along this paper, we only consider
$T_{\rm eff}$ values $\lid \: 40$ kK.  The relations derived for the highest temperatures could be used only to estimate the order of magnitude of the radiation field effective temperature.

\subsection{$Z$  estimations}

To estimate the metallicity $Z$, we use strong-line methods. Nowadays,
several authors have proposed  empirical calibrations of $Z$ with different 
 strong emission lines combinations (see  \citealt{lopezsanchez09} for a review) and  different
calibrations can produce values in disagreement from
each other by until 0.7 dex \citep{kewley08}.

However, since  there is consensus that  the $T_{\rm e}$-method yields 
 more reliable metallicity (or abundance)  estimations, empirical calibrations based on direct determinations of the electron temperature of nebulae \citep{leonid01, leonid00} are preferable  than those relations  
theoretically developed  (see, for example, the calibrations proposed by \citealt{kewley02}). 
Moreover, this kind of  strong emission-line calibrations have an advantage with respect to the others since the physical conditions of the nebulae (established by 
$T_{\rm eff}$, geometry, mass, etc,  essential ingredients in order to estimate the abundance)
are taken into account via the direct determination of the electron 
temperature. 

Therefore, along this paper, we used a strong emission-line calibration to estimate the metallicities. In particular we adopted the empirical calibration proposed by \citet{pilyugin16}
\begin{eqnarray}
  \begin{array}{lll}
12+{\rm (O/H)}_{R,2D} \!\!\!&=&\!\!\!  8.589 + 0.329  \, \log N_{2} \, + \\  
                      &&\!\!\! + (-0.205 + 0.549 \, \log N_{2}) \times \log R_{2},   \\ 
     \end{array}
\label{equation:ohr1p}
\end{eqnarray}

 \noindent where $N_2 = I_{\rm [N\,II] \lambda\lambda 6548+84} /I_{{\rm H}\beta }$
 and $R_2$  = $I_{\rm [O\,II] \lambda\lambda 3726+29} /I_{{\rm H}\beta }$.


\begin{table}
\caption{Coefficients of the linear regression $T_{\rm eff}= {\rm a} \: \times \: R+ {\rm b} $ estimated for two different
ranges of  $T_{\rm eff}$ (in units of $10^{3}$ K), 
where $ R=\log($[\ion{O}{ii}]$\lambda\lambda$3726+29/[\ion{O}{iii}]$\lambda$5007).}
\label{tab1}
\begin{tabular}{@{}lcccc@{}} 
\hline
                                &      \multicolumn{2}{c}{$T_{\rm eff}= 30-40$}     &    \multicolumn{2}{c}{$T_{\rm eff}= 40-50$}  \\
$\log U$                  &                     a                               &          b                 & a     &     b   \\ 
\noalign{\smallskip}				
\hline
                             &                         \multicolumn{4}{c}{$Z=Z_{\odot}$}                                \\
\noalign{\smallskip}
\cline{2-5}
\noalign{\smallskip}
$-$1.5                       &         $-3.69\pm0.10$           &   37.27$\pm0.12$        &   $-33.17\pm2.39$  &  16.85$\pm2.05$  \\  
$-$2.0                       &         $-3.27\pm0.20$           &   38.60$\pm0.31$        &   $-30.00\pm0.60$  &   30.88$\pm0.29$  \\  
$-$2.5                       &         $-2.95\pm0.20$           &   39.78$\pm0.41$        &   $-31.54\pm0.84$  &   43.95$\pm0.09$  \\ 
$-$3.0                       &         $-2.90\pm0.14$           &   41.39$\pm0.37$        &   $-30.43\pm1.71$  &   59.65$\pm0.85$   \\ 
$-$3.5                       &         $-3.20\pm0.10$           &   44.05$\pm0.31$        &   $-26.36\pm2.15$   &  76.17$\pm2.56$   \\   
\hline
                             &                         \multicolumn{4}{c}{$Z=0.5 \:Z_{\odot}$}                                \\
\noalign{\smallskip}
\cline{2-5}
\noalign{\smallskip}
$-$1.5                       &         $-3.02\pm0.13$          &   37.14$\pm0.17$        &  $-42.48\pm2.24$      & $-4.58\pm2.63$   \\  
$-$2.0                       &         $-2.70\pm0.02$          &   38.30$\pm0.05$        &  $-38.98\pm1.76$      &  15.12$\pm1.36$  \\  
$-$2.5                       &         $-2.52\pm0.04$          &   39.48$\pm0.09$        &  $-36.84\pm1.01$      & 34.47$\pm0.30$    \\ 
$-$3.0                       &         $-2.50\pm0.05$          &   40.93$\pm0.16$        &  $-33.52\pm0.54$       &  53.94$\pm0.15$  \\ 
$-$3.5                       &         $-2.73\pm0.08$          &   43.18$\pm0.26$        &  $-29.43\pm0.28$       & 74.10$\pm0.28$    \\   
\noalign{\smallskip}
\hline
                             &                         \multicolumn{4}{c}{$Z=0.2 \:Z_{\odot}$}                                \\
\noalign{\smallskip}
\cline{2-5}
\noalign{\smallskip}
$-$1.5                       &        $-2.44\pm0.24$           &   36.44$\pm0.37$         &   $-33.36\pm7.28$   &  $-1.39\pm10.16$ \\  
$-$2.0                       &        $-2.54\pm0.04$           &   38.21$\pm0.08$         &   $-32.14\pm2.81$   &  15.47$\pm2.60$   \\  
$-$2.5                       &        $-2.40\pm0.03$           &   39.35$\pm0.07$         &   $-31.63\pm0.99$   &  32.52$\pm0.40$   \\ 
$-$3.0                       &        $-2.39\pm0.05$           &   40.82$\pm0.15$         &   $-30.50\pm0.40$   &  51.47$\pm0.09$   \\ 
$-$3.5                       &        $-2.62\pm0.07$           &   43.04$\pm0.26$         &   $-28.69\pm0.23$   &  72.69$\pm0.23$   \\   
\noalign{\smallskip}
\hline
                             &                         \multicolumn{4}{c}{$Z=0.03 \:Z_{\odot}$}                                \\
\noalign{\smallskip}
\cline{2-5}
\noalign{\smallskip}
$-$1.5                       &         $-2.46\pm0.20$             &   37.06$\pm0.33$       &   $-31.45\pm2.04$    &   $-5.31\pm3.28$  \\  
$-$2.0                       &         $-2.38\pm0.05$             &   38.06$\pm0.11$       &   $-29.67\pm1.10$    &   13.42$\pm1.18$  \\  
$-$2.5                       &         $-2.28\pm0.02$             &   39.20$\pm0.06$       &   $-30.58\pm0.59$    &   30.37$\pm0.29$  \\ 
$-$3.0                       &         $-2.30\pm0.05$             &   40.70$\pm0.15$       &   $-30.01\pm0.26$    &   50.33$\pm0.05$  \\ 
$-$3.5                       &         $-2.45\pm0.08$             &   42.72$\pm0.23$       &   $-28.08\pm0.23$    &   72.16$\pm0.23$  \\   
\hline
\end{tabular}
\end{table}



\subsection{$U$ estimations}
\label{ion}

Concerning  calibrations between $U$ and strong emission-lines,  \citet{diaz91}, \citet{dors11}, \citet{sanders16} and \citet{morisset16}, using photoionization
model results, proposed theoretical relations between this parameter and 
[\ion{S}{ii}]($\lambda\lambda$6717+31)/[\ion{S}{iii}]($\lambda\lambda$9069+9532),
 [\ion{O}{ii}]$\lambda\lambda$3726+29/[\ion{O}{iii}]$\lambda$5007 and   [\ion{S}{ii}]($\lambda\lambda6717+31$)/H$\alpha$)
  emission-line ratios.
A large fraction of the variation of $U$ is due to differences in the temperature of the ionizing stars through the amount of the hydrogen 
ionizing photons. 

Among the line ratios above,  we consider a calibration between the ionization parameter $U$ and 
the emission-lines ratio defined as

 \begin{equation}
S2= \rm \log ( [S\:II](\lambda\lambda6717+31)/H\alpha).
\end{equation}

This emission-line ratio is preferable to be used to derive $U$  
due to  its (relatively) low dependence on the $T_{\rm eff}$. In fact, \citet{pellegrini11} used 
 diagnostic diagrams containing several  model-predicted  and observed emission-line ratios in order to study the physical conditions
 and ionization mechanisms across the 30 Doradus \ion{H}{ii} region. By using a large grid of models built with the {\sc Cloudy} code \citep{ferland13}, these authors showed that,
 for a fixed metallicity value, $S2$ has a maximum variation of $\sim 0.2$ dex for  models with $T_{\rm eff}$ ranging from 36 kK to 44 kK.
 (see Fig.5(b) of \citealt{pellegrini11}).
 
\begin{figure}
\centering
\includegraphics[angle=-90,width=1\columnwidth]{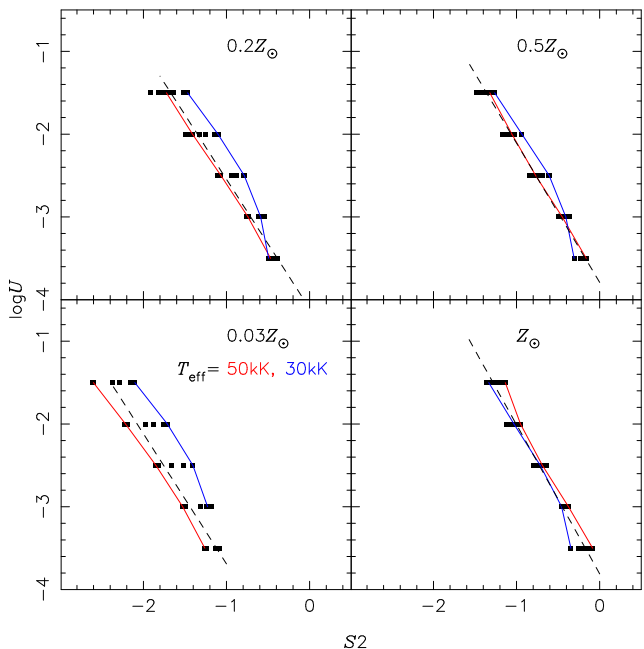}
\caption{Logarithm of the ionization parameter ($\log U$) versus the line ratio  $S2= \rm \log [S\:II](\lambda\lambda6717+31)/H\alpha$
for different values of metallicity ($Z/Z_{\odot}$) and different $T_{\rm eff}$ values, as indicated.
 Points, linked by solid lines, represent the results of our photoionization models
 (see Sect.~\ref{mod}) for $T_{\rm eff}$ values  of 30 kK and 50 kK, as indicated. Dashed lines
 represent linear regression on  the average of $S2$ values for each $\log U$, whose the coefficients
 are listed in Table~\ref{tab1a}.}
\label{f3}
\end{figure}

  In Fig.~\ref{f3}, the results of our models for the relation $\log U$-$S2$  and for different metallicity and
 $T_{\rm eff}$ values are shown. Results for $T_{\rm eff}$ values of 30 kK and 50 kK are linked
 by solid lines. We can see that, for $(Z/Z_{\odot}) \: \gid \: 0.2 $  and for a fixed value of $S2$,
 $\log U$  ranges up to $\sim 0.2$ dex when $T_{\rm eff}$   varies between 30 kK and 50 kK. 
 A higher variation,  up to $~\sim 0.4$ dex, is obtained for $(Z/Z_{\odot}) \: < \: 0.2 $. 
 
 In order to produce  an expression for the $\log U$-$S2$ relationship that be independent of $T_{\rm eff}$, 
 we calculated  an  $S2$ average value for each $\log U$ and did a linear
 regression for  each of the four metallicity values considered. These fitting are represented in Fig.~\ref{f3} 
 by dashed lines and  the coefficients resulting are listed in Table~\ref{tab1a}. It can be seen that there is a dependence of the coefficients ``a" and ``b" with the metallicity. Taken into account this dependence we were able to re-write the $\log U$-$S2$ relationship as:
\begin{equation}
\label{re1}
\log U= {\rm a}(Z/Z_{\odot}) \: \times S2+ {\rm b}(Z/Z_{\odot}), 
\end{equation}
 where
 \begin{equation}
 \label{re2}
 {\rm a}=-0.26(\pm 0.04) \times \: (Z/Z_{\odot}) -1.54 (\pm0.02)
 \end{equation}
 and
\begin{equation}
\label{re3}
\noindent\begin{array}{lcl} 
{\rm b} \!\!\!\!&=&\!\!\!\! -3.69(\pm 1.79) \times (Z/Z_{\odot})^{2} +5.11 (\pm 1.96)  \times (Z/Z_{\odot}) +\\
\!\!\!\!&&\!\!\!\! -5.26 (\pm 0.36). 
\end{array}
\end{equation}

\noindent A similar expression, taking  into account the metallicity dependence on the $\log U$-$S2$, was obtained by \citet{diaz91}.

 \begin{table}
\caption{Coefficients of the  fitting  $\log U= {\rm a} \: \times \: S2+ {\rm b}$,  calculated  for  different
ranges of  $ Z/Z_{\odot}$ and  from the average of $S2$ values for each $\log U$,  
where $ S2=\rm \log [S\:II](\lambda\lambda6717+31)/H\alpha.$ }
\label{tab1a}
\begin{tabular}{@{}lcc@{}} 
\hline
$Z/Z_{\odot}$                  &        a                                   &          b           \\ 
\noalign{\smallskip}				
 0.03                               &      $-1.57 (\pm0.11)$            &     $-5.26 (\pm0.20)$            \\
 0.20                               &      $-1.56 (\pm0.08)$            &     $-4.11 (\pm0.09)$          \\
 0.50                               &      $-1.68 (\pm0.05)$            &     $-3.79 (\pm0.04)$        \\
 1.00                               &      $-1.80 (\pm0.05)$            &     $-3.81 (\pm0.04)$     \\
 \hline
\end{tabular}
\end{table}

\section{Uncertainty in $T_{\rm eff}$ estimations}
\label{err}

The uncertainty in deriving the $T_{\rm eff}$ through  emission-lines
is mainly due to the error in their measurements and the dependence of these lines with 
some nebular parameters (e.g. electron density, metallicity, ionization degree, etc.), the presence 
of multiple ionizing stars within the \ion{H}{ii} region and/or the fact that
the nebular emission could arise from a complex of  \ion{H}{ii} regions
rather than a single region. In what follows, each source of uncertainty is analysed.

\subsection{Nebular parameters and line-measure uncertainties}

\subsubsection{Metallicity}

Abundance determinations of \ion{H}{ii} regions have showed
that these objects exhibit a large range of metallicity, from $Z \approx Z_{\odot}$ for the most metallic objects 
 (e.g. \citealt{dors08, bresolin04, kenniccut03}) 
to $Z \approx Z_{\odot}/30$ for the poorest ones (e.g. \citealt{garnett94, skillman93}). 
The metallicity is one of the key parameters that control the emission-line intensities and, hence, the relative intensity between the lines. 
 Nevertheless, from Fig.~\ref{f1} we can see that the relation $T_{\rm eff}$-$R$ has a weak dependence with the metallicity since the parameter space occupied by the photoionization models with different Z is almost the same. For example, if we  assume   $R$=1 and  
$\log U=-2.5$, $T_{\rm eff}$  varies of only  $\sim$100\,K
for $Z$ values from 0.03 to 1 $Z_{\odot}$.  
 However, as shown
in Fig.~\ref{f1}, the $T_{\rm eff}$-$R$ relation is  strongly dependent on  $U$  which is derived  from $S2$,  which in turn has
a dependence on the metallicity.  In fact, if we assume a fixed value for $S2=-1$ and  
a metallicity   uncertainty of 0.1 dex \citep{pilyugin16}, by using Eqs.~\ref{re1}-\ref{re3}, we found variations in $\log U$ 
of  $\sim0.25$ dex, which  translates into a $T_{\rm eff}$  uncertainty of $\sim 1.0$ kK.

\subsubsection{Ionization parameter}

 Among all nebular parameters, the ionization parameter $U$ is the one 
has the main influence on the $R$-$T_{\rm eff}$ relation. 
 In fact, \citet{oey00} presented a detailed comparison of spectra of spatially resolved
  \ion{H}{ii} regions with photoionization model results. These authors showed that emission
lines of species of ions with low ionization degree, such as the [\ion{S}{ii}]($\lambda\lambda$6717+31),
have a scatter of about 0.3 dex along the objects analysed, indicating a local variation in the ionization
parameter $U$. Moreover, \citet{pellegrini11}, who produce a detailed comparison between model predicted and observed emission line
intensities of 30 Doradus, showed that variations in $U$ of until $~1$ dex can be found along  this object.
Assuming this value, we would have an uncertainty in $U$ of about $0.5$ dex. Taking into account this error in $U$ and assuming $R=1.0$ and $Z/Z_{\odot}=1.0$, a $T_{\rm eff}$ error of $\sim$2.0 kK was estimated from
Fig.~\ref{f1}.
 
\citet{pellegrini11}  also showed the influence of the optical depth on the $S2$ ratio, where
optically thick models are needed to describe low $S2$ values found in some parts
of 30 Doradus.  Moreover, these authors also pointed out that many lower metallicity nebular regions in 30 Doradus have
 $S2$ affected by density, non radiation bound. Obviously, these process
affects the use of $S2$ as a tracer of $U$.

However, it is worth noting that the values of the ionization parameter  based on our  $U$-$S2$ calibration (Eq.~\ref{re1}) must be
interpreted as a global nebular parameter since it is derived from observations and models of the integrated flux of the objecs, and this equation can not be used
for spatially resolved studies as the one done by \citet{pellegrini11} and \citet{oey00}.
In any case, assuming that $U$ is correct by 0.2 dex (see Sect.~\ref{ion}),  $R=1.0$  and $Z/Z_{\odot}=1.0$, 
from Fig.~\ref{f1}, we found  a $T_{\rm eff}$ uncertainty of $\sim 1.0$ kK.

\subsubsection{Emission line errors}

Regarding the uncertainty in the emission-line flux measurements,
typical errors for the strong emission-lines involved in our relations are between about 1 and 5\% (e.g.\  \citealt{hagele08, hagele06, kenniccut03}).
 Assuming that the estimation of the 
$N2$, $R2$ and $S2$  line ratios have errors as high as 6\%, it yields 
an error in $Z$ of $\sim$3\% and in $U$ of $\sim$0.1 dex. Taking into account this $U$
uncertainty and also considering  an error of 6\% for the $R$ index,  
we found a $T_{\rm eff}$ error of about $\sim0.5$ kK due to  errors in the emission line measurements.

 \subsection{Multiple stars presence}

It is known that giant star-forming regions   are ionized
by multiple stars, i.e. by an ionizing  cluster containing 
stars with a large range of mass and evolutive stages (e.g. \citealt{pellegrini10, bosch01, mayya96}).
Thus, the derivation of an unique $T_{\rm eff}$ values
for these cases could be somewhat  uncertain.

To test the effect of the presence of multiple stars in our
$T_{\rm eff}$ estimations, we performed a simple analysis.
Firstly, we calculated a photoionization model with solar metallicity and
ionized by a  single star with spectral type O7V. 
Assuming the calibrations presented by \citet{massey11} and \citet{dekoter97}, 
this star has a mass of about $M$=30 $M_{\odot}$,
$T_{\rm eff}$=37 kK and luminosity $\log(L/L_{\odot})=6$.
 This model predicts the line ratios ($R$, $N2$, $R2$, $S2$)=(1.31, 0.93, 1.77,$-0.27$), for which, using 
the methodology presented in Section~\ref{met}, we derived $T_{\rm eff}\approx \,37.65$ kK.

Secondly, we assumed a photoionization model with the same parameters than the
previous one, but having  as the ionizing source 
a stellar cluster, whose
spectral energy distribution was obtained from the {\sc STARBURST99}  synthesis code \citep{leitherer99}. 
We built a synthetic spectrum considering an instantaneous-burst stellar cluster, a Salpeter initial mass function \citep{salpeter55} 
with an upper mass limit of $M$=30 $M_{\odot}$, an age of 2.5 Myr, 
a solar metallicity, and the same atmosphere models than the ones used in the models presented in Section~\ref{mod}.
According to \citet{leitherer99}, this stellar cluster has a
luminosity of $\log(L/L_{\odot})=8.72$.
 We found that this model predicts ($R$, $N2$, $R2$, $S2$)=(1.52, 0.37, 1.74,$-0.80$), 
which also indicates $T_{\rm eff}\approx35.65$\,kK.

Thus, we showed that, for giant \ion{H}{ii} regions generally 
ionized by a stellar cluster, the  assumption
of a  single star as the main ionizing source of the
gas is correct.
 Therefore, we assume an uncertainty of  2 kK in
$T_{\rm eff}$ estimations due to multiple stars presence.


\subsection{\ion{H}{ii} region complex}

In general, for distant objects, the observed spectra comprise the flux of a complex of \ion{H}{ii} regions and the physical properties derived represent an averaged value (e.g. \citealt{rosa14, krabbe14, hagele13, hagele10, hagele09, hagele07}). 
 In principle, this is not critical in our $T_{\rm eff}$ estimations because \ion{H}{ii} regions comprising the complexes were probably formed from a same parent molecular cloud with similar initial conditions and resulting in similar stellar contents and nebular parameters (see e.g. \citealt{kenniccut03}). 
 
To simulate the $T_{\rm eff}$ estimations in an \ion{H}{ii} region complex, we use as prototype system two pairs of near \ion{H}{ii} regions located in the spiral galaxy NGC\,2403 and observed by \citet{garnett97}. These are the pairs VS35-VS24 (refereed as C1) and VS49-VS48 (refereed as C2) located at about 1 kpc and 5.6 kpc from the NGC\,2403 centre, respectively. Initially, considering the observational data obtained by \citet{garnett97}, we derived one $T_{\rm eff}$ value for each individual \ion{H}{ii} region of each pair. VS35 and VS24 have the emission-line  intensity ratios ($R$, $N2$, $R2$, $S2$) equal to (0.25, 0.45, 2.46, $-0.32$) and (0.28, 0.43, 2.41, $-0.46$), that following the methodology presented in Sect.~\ref{met}, translate into the
$T_{\rm eff}$ values of 37.6 kK and 37.5 kK, respectively.
Now, adding the emission line fluxes of individual objects VS35 and VS24   we found for C1 $T_{\rm eff}=37.5$. The same procedure was considered for the VS49 and VS48 and we found about the same $T_{\rm eff}$ values for these individual objects and for C2, i.e. $~40$ kK. Thus, we show that our method produces an averaged $T_{\rm eff}$ values in \ion{H}{ii} region complexes and no uncertainties is yielded for distant objects.

\bigskip

 Along the paper, we will consider that the $T_{\rm eff}$ estimation is correct by 2.5 kK, the quadratic sum of the uncertainties discussed above.

 
\section{Results}
\label{res}

\begin{figure}
\centering
\includegraphics[angle=-90,width=1\columnwidth]{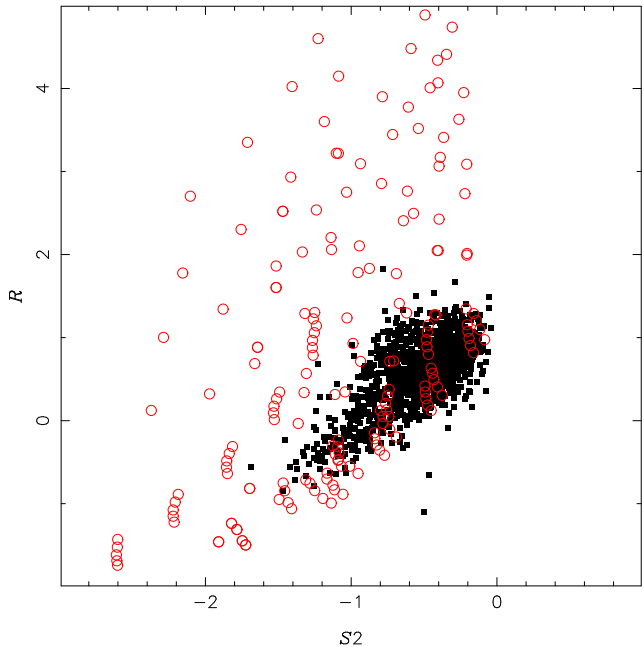}
\caption{$R$=log([\ion{O}{ii}]$\lambda\lambda$3726+29/[\ion{O}{iii}]$\lambda$5007)  vs.  $S2$=log([\ion{S}{ii}]($\lambda \lambda$6717+31)/H$\alpha$).
 Black squares represent the observational data (see Sect.~\ref{obs}) while red circles the results of our models (see Sect.~\ref{mod}).}
\label{ana1}
\end{figure}

 In Fig.~\ref{ana1}, the line ratios  $R$ as a function of  $S2$  
predicted by our models are compared with those of the observational sample.  We can see that  our  photoionization models describe very well
the region occupied by the observational data, indicating that the models are representative of real \ion{H}{ii} regions.

In order to eliminate any bias in our analysis that could be yielded by the extrapolation of our models out to the parameter space sampled by them, 
we only consider those objects listed in Table~\ref{tab2} whose estimated metallicities, ionization parameters and $T_{\rm eff}$ are in the ranges sampled by 
our grid of photoionization models: $0.03 \: \lid \: Z/Z_{\odot} \lid \: 1.0$,
$-3.5 \: \lid \: \log U \: \lid \:-1.5$ and $30 \: \lid \: T_{\rm eff} (\rm kK)  \: \lid  \: 40$.
This make possible to estimate $T_{\rm eff}$ values  for 865 ($\sim$72\% of the sample) \ion{H}{ii} regions of our sample.

\begin{figure}
\centering
\includegraphics[angle=-90,width=1\columnwidth]{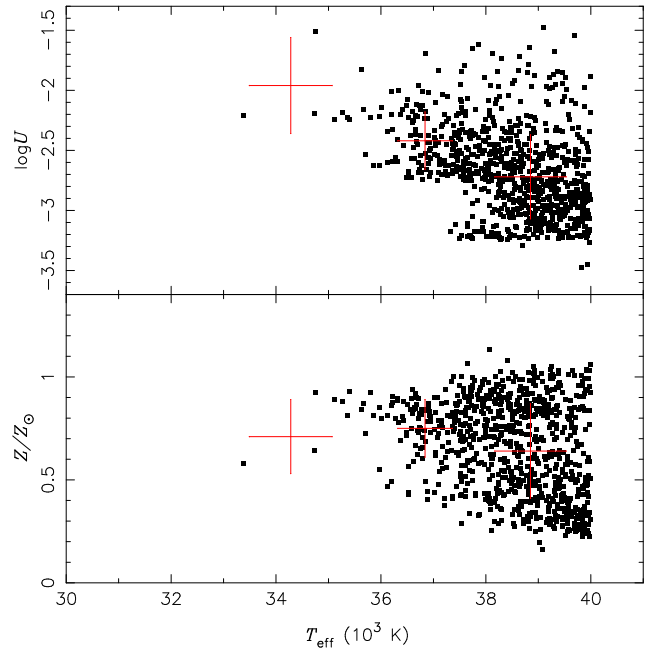}
\caption{$Z/Z_{\odot}$ and $\log U$ as a function of $T_{\rm eff}$. Points represents estimations for the sample presented 
in Table~\ref{tab2} for which were possible to compute the parameters considered.
Error bars represent the standard deviation of the average values presented in Table~\ref{tab1a}.
The error in $T_{\rm eff}$ was assumed to be 2\,000 K (see Sect.~\ref{err}).}
\label{f6}
\end{figure}

In Fig.~\ref{f6}, $Z/Z_{\odot}$ (lower panel) calculated using the Eq.\,\ref{equation:ohr1p} and $\log U$ (upper panel) calculated using the Eqs.\,\ref{re1}, \ref{re2} and \ref{re3},  as a function of 
the effective temperature $T_{\rm eff}$
for the   865 objects are presented. We can see that for
most of the objects $T_{\rm eff}$ is higher than $36$\,kK,
with an average value of $38.5 (\pm 1.0)$\,kK. Also in Fig.~\ref{f6},
the average and the standard deviation of these parameters
 considering different $T_{\rm eff}$ ranges  (see Table \ref{tab1a4}) are shown.
 We can note that, despite the large dispersion, it seems that an anti-correlation between
$T_{\rm eff}$ and   $\log U$ is obtained.  In contrast, no trend is found between $T_{\rm eff}$ and $Z/Z_{\odot}$.

\begin{table}
\caption{Average values of $T_{\rm eff}$, $Z/Z_{\odot}$,
and $\log(U)$, for the selected ranges of $T_{\rm eff}$. The number 
of objects (N) used  in these calculations are listed.}
\vspace{0.3cm}
\label{tab1a4}
\begin{tabular}{@{}lcccc@{}}
\hline		 
\noalign{\smallskip}  
Range ($10^{3}$ K)  &   $<T_{\rm eff}>$&  $<Z/Z_{\odot}>$     & $<\log U >$                 & N\\
\noalign{\smallskip}   
30.0-32.5           &       ---                            &      ---                      &  ---                        & 0 \\
32.5-35.0           &      $34.28\pm0.79$        &      $0.71\pm0.18$    &  $-1.96\pm0.40$     & 3 \\
35.0-37.5           &      $36.84\pm0.52$        &      $0.75\pm0.14$    &  $-2.42\pm0.24$     & 140\\
37.5-40.0           &      $38.85\pm0.68$        &      $0.64\pm0.23$    &  $-2.72\pm0.35$     &  722\\
\hline
\end{tabular}
\end{table}

\begin{figure*}
\centering
\includegraphics[angle=-90,width=0.6\columnwidth]{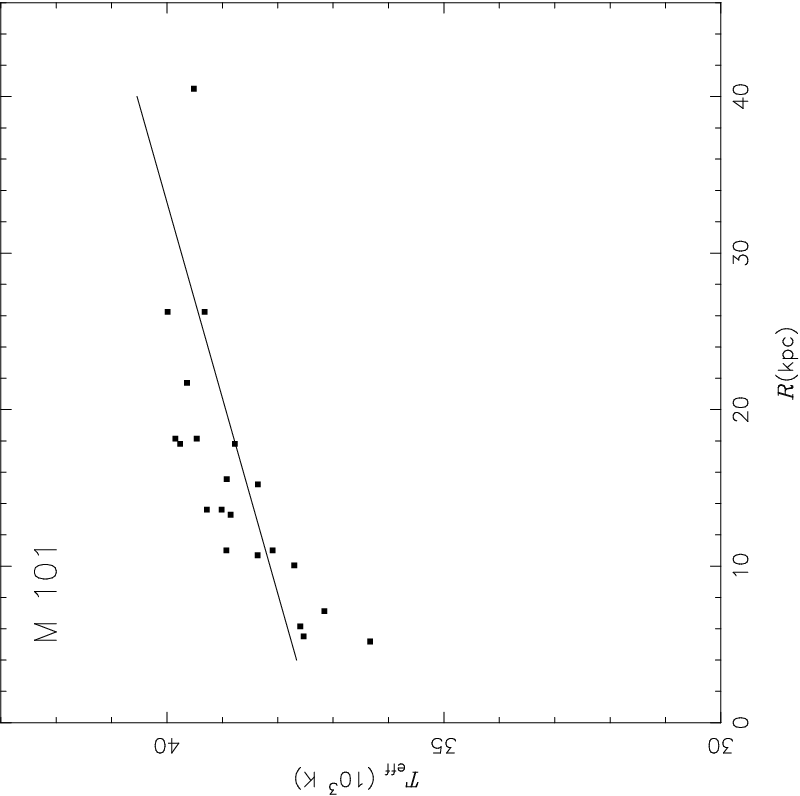}\hspace{0.4cm}
\includegraphics[angle=-90,width=0.6\columnwidth]{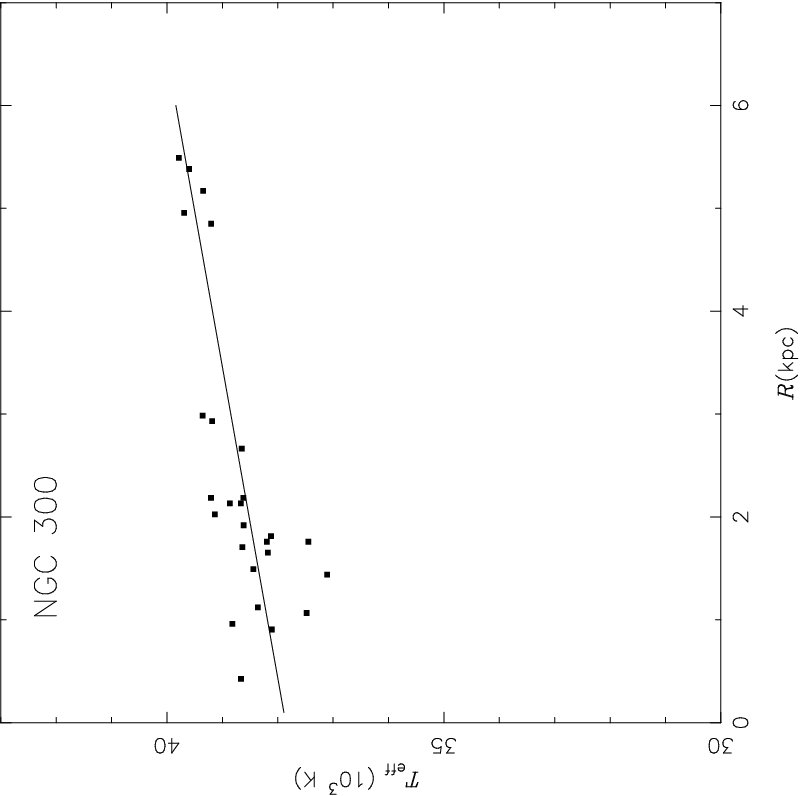}\\\vspace{0.4cm}
\includegraphics[angle=-90,width=0.6\columnwidth]{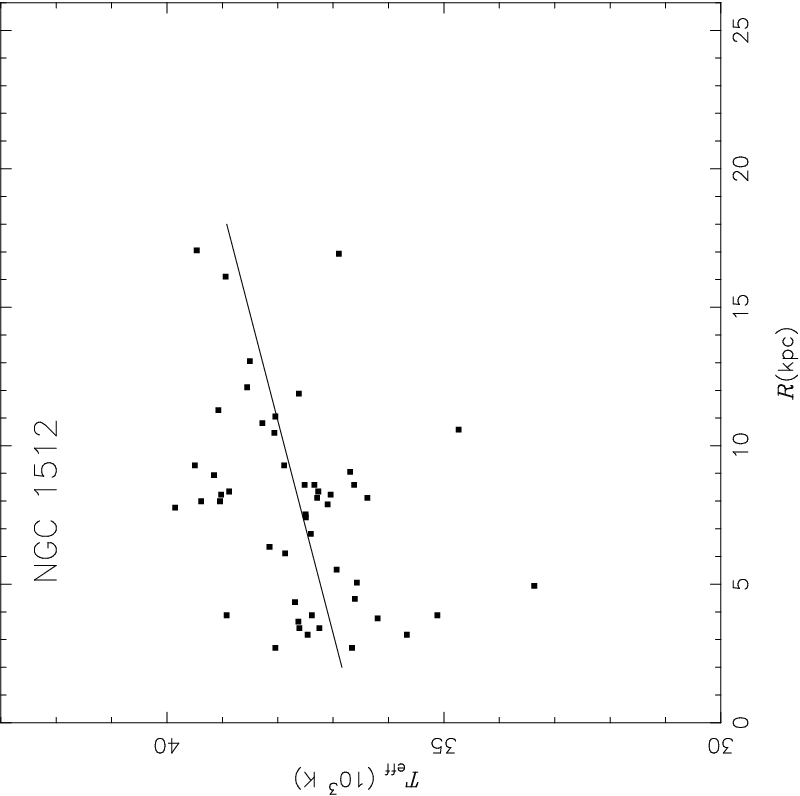}\hspace{0.4cm}
\includegraphics[angle=-90,width=0.6\columnwidth]{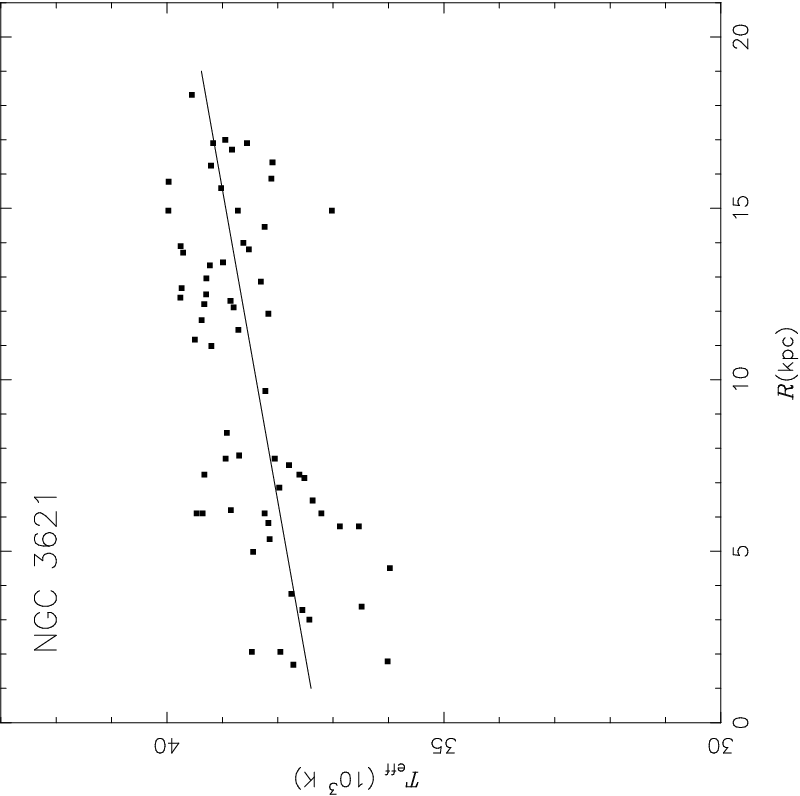}\\\vspace{0.4cm}
\includegraphics[angle=-90,width=0.6\columnwidth]{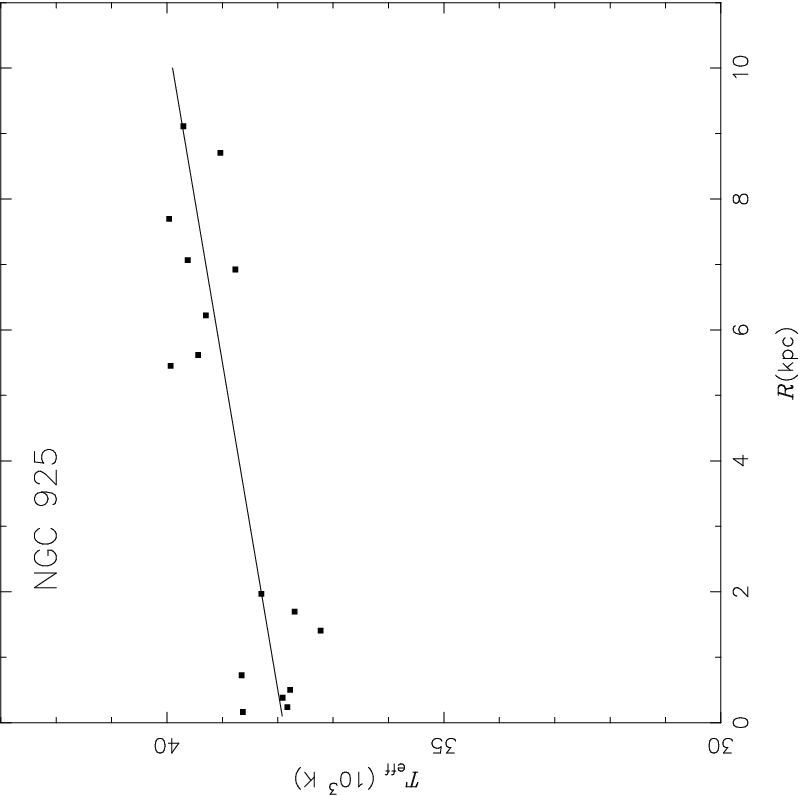}\vspace{0.4cm}
\includegraphics[angle=-90,width=0.6\columnwidth]{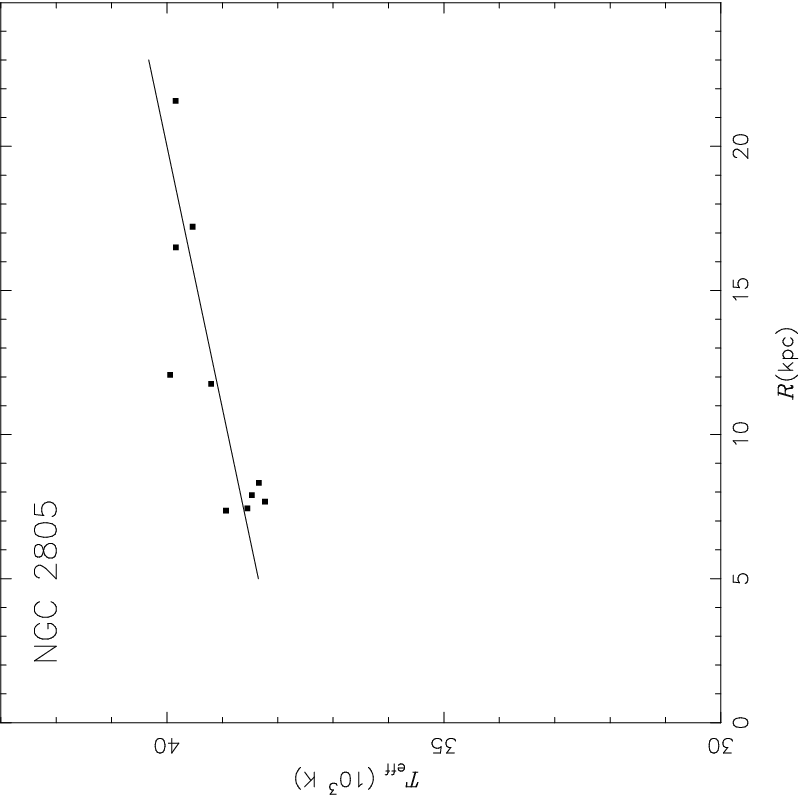}\vspace{0.4cm}
\caption{$T_{\rm eff}$ as a function of the galactocentric distance $R$ for the indicated spiral galaxies. Solid lines represent the linear
regressions whose coefficients are listed in Table~\ref{tab3}. 
The assumed $T_{\rm eff}$ error was 1\,500 K (see Sect.~\ref{err}). }
\label{f7}
\end{figure*}

\begin{figure*}
\centering
\includegraphics[angle=-90,width=0.6\columnwidth]{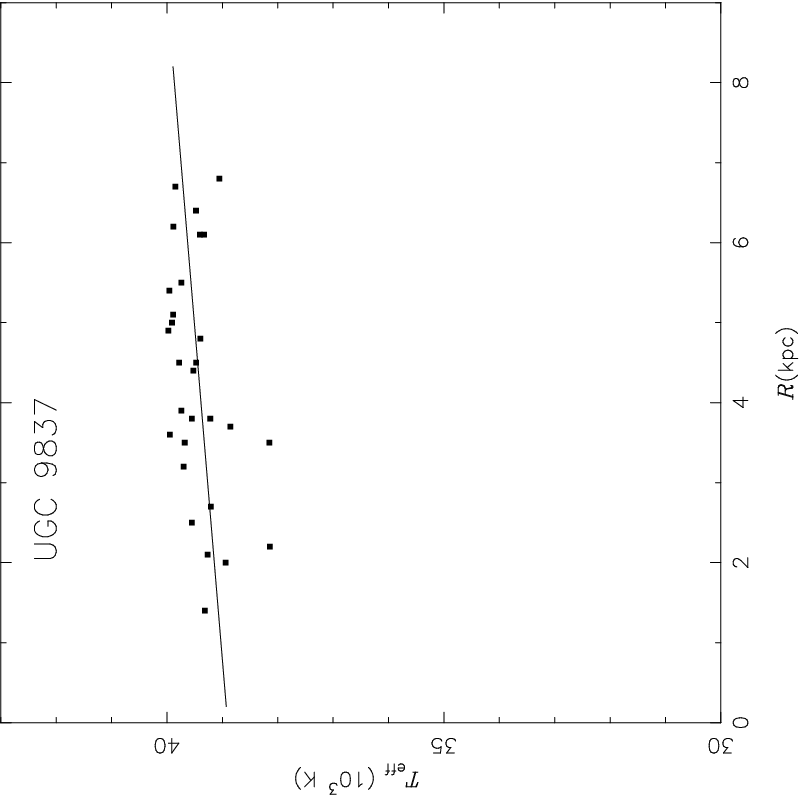}\hspace{0.4cm}
\includegraphics[angle=-90,width=0.6\columnwidth]{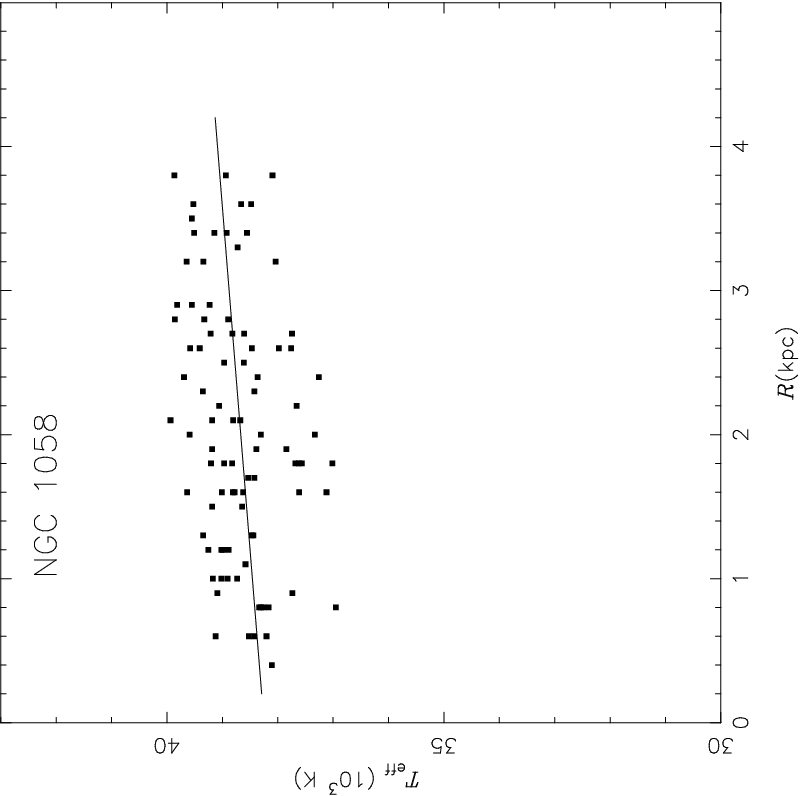}\\\vspace{0.4cm}
\includegraphics[angle=-90,width=0.6\columnwidth]{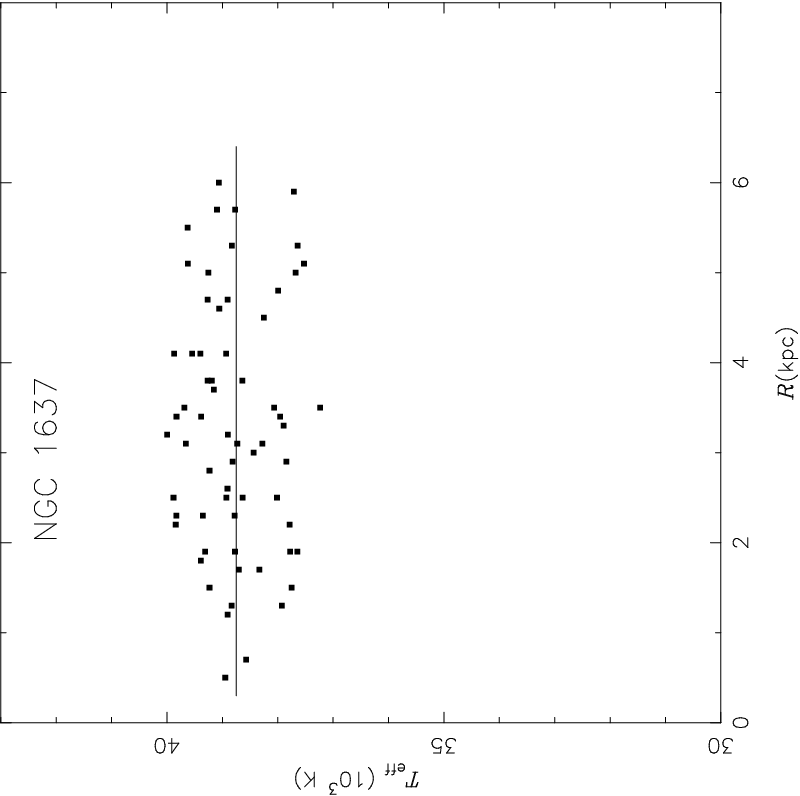}\hspace{0.4cm}
\includegraphics[angle=-90,width=0.6\columnwidth]{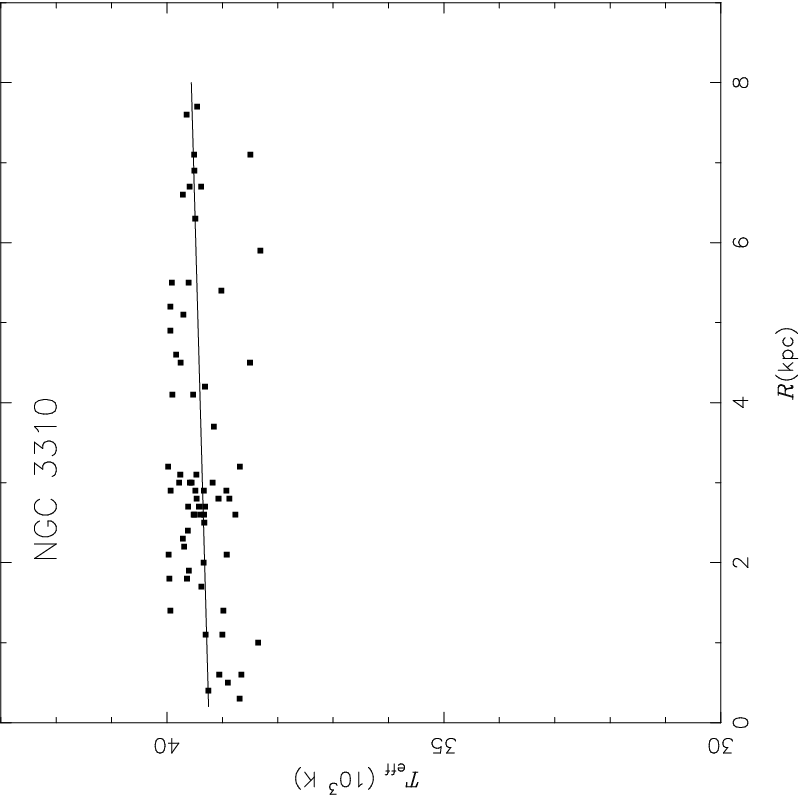}\\\vspace{0.4cm}
\includegraphics[angle=-90,width=0.6\columnwidth]{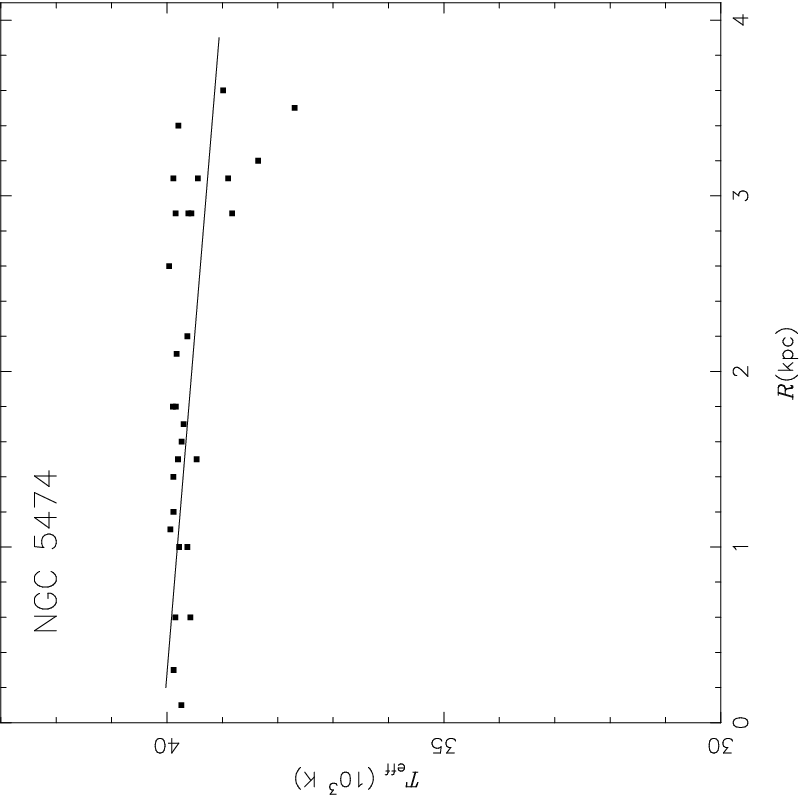}\vspace{0.4cm}
\includegraphics[angle=-90,width=0.6\columnwidth]{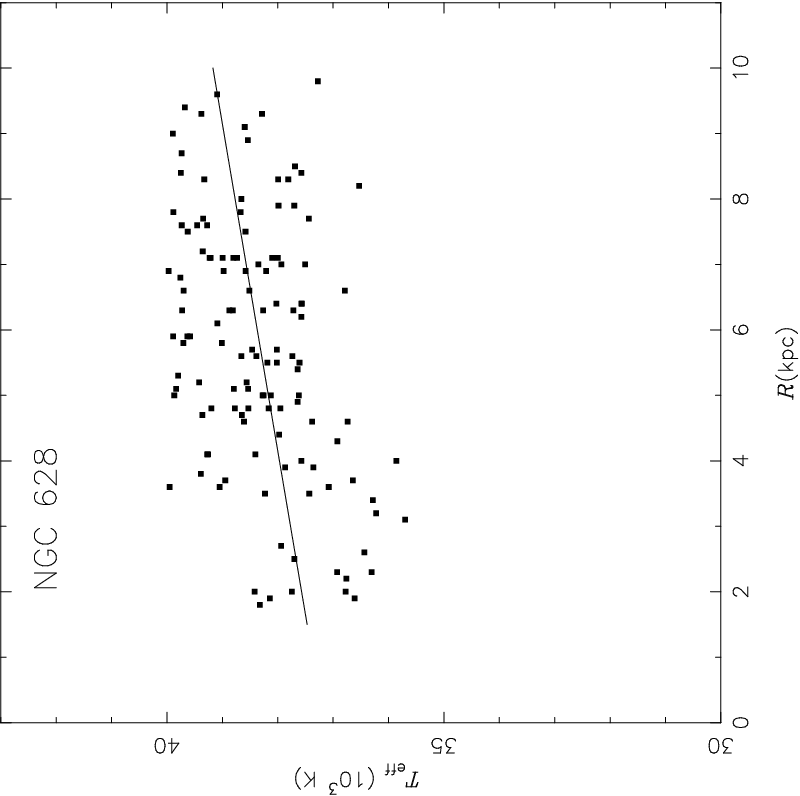}\vspace{0.4cm}
\caption{Same as Fig.~\ref{f7}.}
\label{f8}
\end{figure*}

In our sample there are 14 spiral galaxies for which there are emission lines measurements in least 10 \ion{H}{ii} regions distributed along their disks. Using these observations we estimate the $T_{\rm eff}$ of each \ion{H}{ii} region to investigate the behaviour of the $T_{\rm eff}$ as a function of the galactocentric distances $R$. In Table~\ref{tab3} the coefficients of the linear regressions are listed.
These linear regressions together with the estimated $T_{\rm eff}$  for each \ion{H}{ii} region are plotted in Figs.~\ref{f7} and \ref{f8} as a function of $R$ (in kpc).
For 11 galaxies we  derive a positive  slope,  two objects  (NGC\,1647  and NGC\,3184)  present
a null  slope and only one galaxy (NGC\,5474) shows a negative slope (see Table~\ref{tab3}).

\begin{table}
\caption{Coefficients of the linear regression  $T_{\rm eff} {\rm (kK)}= {\rm a} \: \times \: R ({\rm kpc})+ {\rm b} $
and the number of  \ion{H}{ii} regions used for each galaxy.}
\label{tab3}
\begin{tabular}{@{}lccc@{}} 
\noalign{\smallskip}
\hline
Galaxy            &            a                      & b                        & Number \\
\noalign{\smallskip}
M\,101             & $+0.08(\pm0.02)$       &  $37.34(\pm0.32)$   & 21  \\
NGC\,300        &  $+0.33(\pm0.06)$      &  $37.86(\pm0.18)$    & 26  \\
NGC\,1512      &  $+0.13(\pm0.05)$       &  $36.58(\pm0.41)$   & 49  \\
NGC\,3621      &  $+0.11(\pm0.02)$      &   $37.29(\pm0.29)$   & 63   \\
NGC\,925        &  $+0.20(\pm0.04)$       &  $37.90(\pm0.21)$   & 16   \\
NGC\,2805      &  $+0.11(\pm0.02)$       &  $37.80(\pm0.34)$    & 12   \\
UGC\,9837      &  $+0.12(\pm0.05)$       &  $38.91(\pm0.25)$   & 29   \\
NGC\,1058      &   $+0.21(\pm0.07)$     &  $38.25(\pm2.54)$    &  91  \\
NGC\,1637      &   $0.00(\pm0.06)$      &   $38.75(\pm0.23)$   & 64   \\
NGC\,3310      &   $+0.04(\pm0.02)$    &   $39.24(\pm0.10)$   & 67  \\
NGC\,5474      &   $-0.26(\pm0.08) $    &  $40.08(\pm0.18)$    &  29   \\
NGC\,628        &   $+0.20(\pm0.04)$    &   $37.17(\pm0.25)$   &  125   \\
NGC\,1232      &   $+0.16(\pm0.02)$    &   $35.79(\pm0.31)$   &  16   \\
NGC\,3184      &   $0.00(\pm0.10)$    &   $37.53(\pm0.56)$   &  17   \\
\hline
\end{tabular}
\end{table}

\begin{figure*}
\centering
\includegraphics[angle=-90,width=0.6\columnwidth]{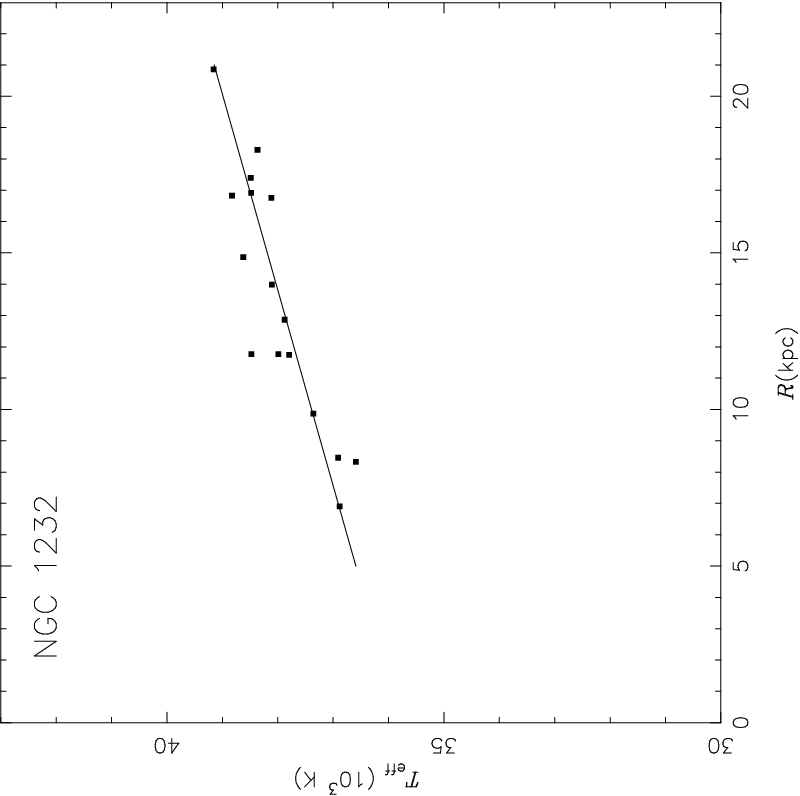}\hspace{0.4cm}
\includegraphics[angle=-90,width=0.6\columnwidth]{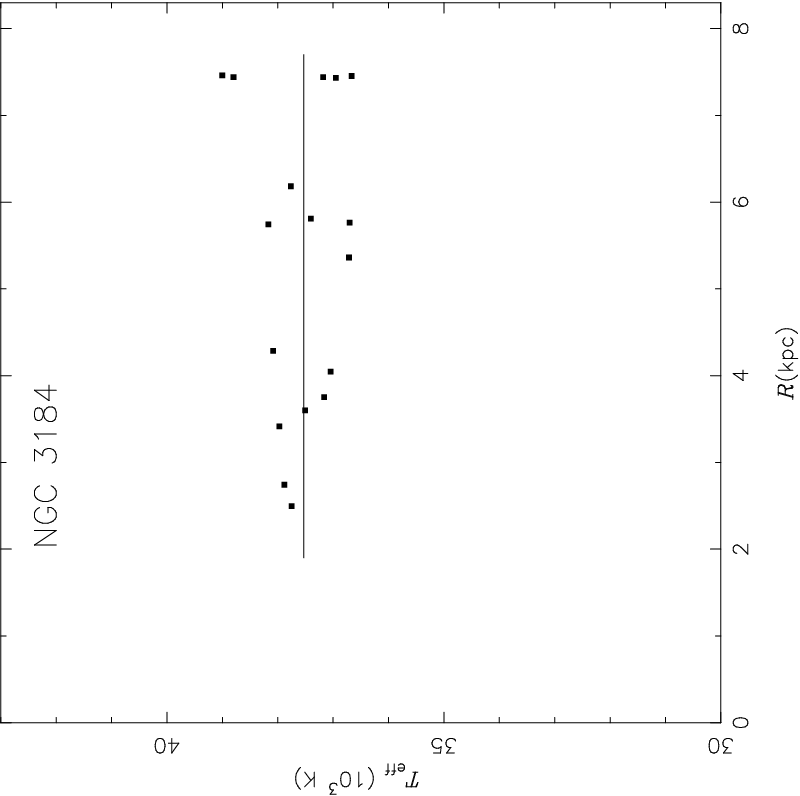}\hspace{0.4cm}
\caption{Same as Fig.~\ref{f7}.}
\label{f9}
\end{figure*}

 
\section{Discussion}
\label{disc}

In the present work, we  propose a calibration between the $R=\log($[\ion{O}{ii}]$\lambda\lambda$3726+29/[\ion{O}{iii}]$\lambda$5007) index
and $T_{\rm eff}$  based on photoionization models built assuming a match between the metalliciy of the gas phase of the hypothetical \ion{H}{ii} 
region and the one of the atmosphere of the ionizing stars.  We took into account the dependence of the $T_{\rm eff}$-$R$ relation  with the metallicity and  with the ionizing parameter. 
Other emission-line ratios have also been proposed  in the literature as $T_{\rm eff}$ indicators. For example, \citet{kennicutt00} used a grid of photoionization models
and  showed that the line ratios  He\,I$\lambda$5876/H$\beta$, [\ion{O}{iii}]$\lambda$5007/H$\beta$,     
among others, can be used to estimate  $T_{\rm eff}$. Also, \citet{oey00} presented a detailed comparison of optical \ion{H}{ii} region spectra with photoionization models
in order to investigate the reliability of some line ratios as $T_{\rm eff}$ indicators. These authors found that, among several line ratios considered, 
the [\ion{Ne}{iii}]$\lambda$3869/H$\beta$ has higher sensitivity to $T_{\rm eff}$ and it is independent of morphology, and is insensitive to gas  shocks, although it is abundance dependent.

As in the case of the $R$ index, $T_{\rm eff}$ estimations based on other  emission-line ratios require the previous determination of the metallicity $Z$ (or abundance) and of the ionization parameter $U$ \citep{oey00}.
In the present work, we propose that, to infer the $T_{\rm eff}$,  the metallicity can be derived through a calibration based on the $T_{\rm e}$-method and $U$  via a calibration between $S2$ and $U$ which depends on $Z$. Finally $T_{\rm eff}$ is obtained from its relation with the $R$ index.

To test the reliability of the results found in this work, we use the observational data of \ion{H}{ii} regions
located in the Large Magellanic Cloud obtained by \citet{zanstrow13}  to estimate their $T_{\rm eff}$  by using the methodology presented in Sect.~\ref{met}.
The obtained $T_{\rm eff}$ were compared with those derived by Zastrow and collaborators, who built detailed photoionization models in order to reproduce the emission line intensities of the individual objects in their sample. 

In Table~\ref{tab4a} we listed the $T_{\rm eff}$ estimations for the objects for which we were able to apply our methodology together with the $T_{\rm eff}$ values from  \citet{zanstrow13}. The difference between both estimation are also listed. We considered the Zastrow et al. estimations obtained assuming the same atmosphere models used by us (WM-Basic models). 
We found that our $T_{\rm eff}$ values are systematically higher with an average difference of 1.45 kK, which is lower than our estimated uncertainties for the  $T_{\rm eff}$ obtained through the $R$ index (2.5 kK).

\begin{table}
\caption{$T_{\rm eff}$ values estimated from the $R$ index and those from detailed photoionization models
by \citet{zanstrow13} for a sample of \ion{H}{ii} regions in the LMC. }
\label{tab4a}
\begin{tabular}{@{}lccc@{}} 
\hline
\noalign{\smallskip}
                                 &                         \multicolumn{2}{c}{$T_{\rm eff}$ (kK)}                      \\
\cline{2-3}
Object                       &              This paper          &     Zastrow et al.                    &  Difference  (kK) \\
\noalign{\smallskip}     
  L\,32                       &              36.75                  &           34.00                           &       2.70           \\
 L\,35                        &              34.93                  &           31.00                           &       3.93           \\
 L\,52                        &              39.39                  &           38.75                           &       0.64           \\
 L\,344                      &              39.48                  &           39.0                             &       0.48           \\
 L\,390                      &              37.21                  &           37.0                             &       0.21           \\
 L\,394                      &              39.75                  &           39.0                             &       0.75           \\
 \hline
\end{tabular}
\end{table}

As can be seen in Fig.~\ref{f6}, where the $T_{\rm eff}$  estimations  for part of our compiled sample (865 objects) are shown,
we found that there is no correlation  between $T_{\rm eff}$ and $Z$, at least for the $T_{\rm eff}$ values in the range
of validity of the proposed $T_{\rm eff}$-$R$ relationship  ($30  \: \la \: T_{\rm eff} ({\rm kK}) \: \la \:40 $). 
This result is in consonance with the one found by  \citet{morisset04},
who  comparing mid-infrared emission-line intensities of Galactic \ion{H}{ii} regions with photoionization model results did  
not find  evidences of any correlation between $T_{\rm eff}$ and $Z$.
\citet{morisset04} showed that, not taking properly into account the effect of metallicity on the ionizing shape of the stellar atmosphere,  would lead to
an apparent decrease of $T_{\rm eff}$ with $Z$.

Regarding the $T_{\rm eff}$ variation along the disk of spiral galaxies, in  a pioneer work, \citet{shields76} interpreted that the enhancement of the equivalent width of the H$\beta$ emission line of a sample of \ion{H}{ii} regions located in the spiral galaxy M\,101, could be due to an increment of the temperature of the hottest exciting stars, implying higher $T_{\rm eff}$ of the radiation field. 
These authors also concluded that high metallicity \ion{H}{ii} regions have lower $T_{\rm eff}$ values than those with low metallicities. 
Other authors (\citealt{dors03,vilchez88,henry95,dors05}) have derived  similar results. 
This  behaviour has been interpreted as being due to effects  on the opacity of the stellar atmospheres rather than differences in the stellar masses  (see \citealt{bresolin99}). In Figs.~\ref{f7}-\ref{f9} we found positive 
$T_{\rm eff}$ gradients for  11 spiral galaxies,  in agreement with the original idea by \citet{shields76}. 
Since our $T_{\rm eff}$-$R$ relationship is only valid for $30 {\rm kK} \: \la \: T_{\rm eff} \: \la \:40 {\rm kK}$,
  slopes of the $T_{\rm eff}$ gradients in spiral galaxies could be higher 
 than the ones  listed in Table~\ref{tab3}. 
\citet{dors05} used photoionization models in order to reproduce the observed gradients of emission-line ratios for  \ion{H}{ii}  regions
located in the normal spiral galaxy M\,101 and in three barred spiral galaxies, namely NGC\,1365, NGC\,925, and NGC\,1073.
These authors derived positive $T_{\rm eff}$ gradients in the range  0.2-0.4  kK/kpc,  with $T_{\rm eff}$ values  up to 50 kK for the outermost regions of the disks of the galaxies analysed.

\citet{fierro86} compared  observational emission-line intensities of \ion{H}{ii} regions  located between 1 and 5 kpc from the
centre of the spiral galaxy NGC\,2403 with those predicted
by a grid of photoionization models by \citet{stasinska82}. \citet{fierro86} found that models assuming $T_{\rm eff}$=35\,kK
are able to reproduce the observational data (see also \citealt{evans86}).
\citet{enrique09a} used the $T_{\rm eff}$  sensitive parameter $\eta'$ defined  by pairs of consecutive ionization stages of 
the same species and introduced by 
\citealt{vilchez88}: \[\eta'=\frac{I([\ion{O}{ii}]\lambda\lambda3726+29\AA)/I([\ion{O}{iii}]\lambda\lambda4959+5007\AA)}{(I([\ion{S}{ii}]\lambda\lambda6717+31\AA)/I([\ion{S}{iii}]\lambda\lambda9069+9532\AA)}.\]
They plotted this parameter as a function of  
the galactocentric distance for ten galaxies obtaining slopes ranging from $0.00 \pm 0.01$ to $-0.11 \pm 0.05$. 
Taking into account that $T_{\rm eff}$ increases as $\eta'$ decreases, this result is in agreement with our own. 
Nevertheless, the study of the $T_{\rm eff}$ behaviour along the disks of spiral galaxies still seems to be an open question in astronomy.

 
\section{Conclusions}
\label{conc}
 
We have proposed a calibration  for the effective temperatures of the radiation field of the ionizing star clusters of \ion{H}{ii} regions 
($T_{\rm eff}$) as a function of the $R$=log([\ion{O}{ii}]$\lambda\lambda$3726+29/[\ion{O}{iii}]$\lambda$5007)
index. This calibration is based on photoionization models
assuming a match between the  metallicity of both ionizing  stars and the gas phase of the \ion{H}{ii}  regions 
as well as considering the effect of the ionizing parameter on the  $R$ index. 
Since the $R$ index shows small variations for $T_{\rm eff}$ values larger than 40\,kK our method is valid in the range sampled by our models with metallicities ($Z/Z_\odot$) between 0.03 and 1, logarithm of the ionization parameter ($\log U$) between $-3.5$ and $-1.5$ and the effective temperature ($T_{\rm eff}$) between 30 and 40\,kK.
We found that this $T_{\rm eff}$-$R$ relation has 
a strong dependence with the ionizing parameter while it shows a weak  direct dependence with the metallicity (variations in $Z$ translate into variations in $U$).

On the other hand taking advantage of that the  [\ion{S}{ii}]($\lambda\lambda$6717+31)/H$\alpha$ emission-line ratio is about 
constant for a large range of $T_{\rm eff}$, we calculated linear regressions between the ionization parameter and this line-ratio for 
the results of our models and for the different metallicity regimes considered.
A large fraction of the variation of the ionization parameter is due to differences in the temperature of the ionizing stars through the amount of the hydrogen ionizing photons. Hence, the use of this particular line-ratio to derive the ionization parameter is preferable over others in the literature due to  its (relatively) low dependence on the $T_{\rm eff}$.

In this work, we explored the different sources of uncertainties in the $T_{\rm eff}$ estimations finding that the main contribution comes from the probable  multiple star presence as the ionizing source of the giant extragalactic \ion{H}{ii} regions rather than a single star. Small contributions comes from uncertainties in the ionization parameter and metallicity estimations, and from the emission-line measurements.

From the $T_{\rm eff}$ estimations for a sample of  865 \ion{H}{ii} regions, we did not find any correlation  between $T_{\rm eff}$  and the metallicity.
We found that  most of the objects ($\sim 87$\%) present $T_{\rm eff}$ values in the range between 37 and 40\,kK. 
Studying the $T_{\rm eff}$ gradients across the disks of 14 spiral galaxies through the use of the estimated $T_{\rm eff}$ of their \ion{H}{ii} regions we found that 11 of them have positive gradients, other 2 present flat gradients, and only one shows a negative gradient. 
Our results supports the original idea by \cite{shields76} that there is a positive gradient of $T_{\rm eff}$ across the disks of spiral galaxies  traced by \ion{H}{ii} regions although more work on this topic is required to confirm this behaviour and to explain the presence of some flat and negative gradients.

\section*{Acknowledgments}
We are very grateful to Christophe Morisset for his  useful comments
and suggestions that helped us to   improve our work. 
We are also grateful to the anonymous referee for his/her very useful comments and suggestions that helped us to substantially clarify and improve our work.
This research has made use of the NASA/IPAC Extragalactic Database (NED) 
which is operated by the Jet Propulsion Laboratory, California Institute of Technology, under contract with the National Aeronautics and Space Administration.

\label{lastpage}


\begin{thebibliography}{99}
\bibitem[Allende-Prieto et al.(2001)]{alendeprieto} Allende-Prieto C., Lambert D.~L., Asplund M., 2001, ApJ, 556, L63
\bibitem[Ambarzumian(1932)]{ambarzumian32} Ambarzumian  V.~A., 1932, Nature, 129, 725
\bibitem[Baldwin et al.(1981)]{baldwin81} Baldwin  J.~A.,  Phillips M.~M., Terlevich R. 1981, PASP, 93, 5
\bibitem[Berg et al.(2013)]{berg13} Berg D.~A. et al., 2013, ApJ, 775, 128
\bibitem[Bosch et al.(2001)]{bosch01} Bosch G., Selman F., Melnick J.,  Terlevich R., 2001, A\&A, 380, 13
\bibitem[Bresolin et al.(1999)]{bresolin99} Bresolin F.,  Kennicutt R.~C.,  Garnett D.~R., 1999, ApJ, 510, 104
\bibitem[Bresolin et al.(2004)]{bresolin04} Bresolin F., Garnett D.~R., Kennicutt R.~C., 2004, ApJ, 615, 228
\bibitem[Bresolin et al.(2009a)]{bresolin09a} Bresolin F. et al., 2009a, ApJ, 700, 309
\bibitem[Bresolin et al.(2012)]{bresolin12} Bresolin F.,  Kennicutt R.~C., Ryan-Weber E., 2012, ApJ, 750, 122
\bibitem[Bresolin et al.(2009b)]{bresolin09b} Bresolin F., Ryan-Weber  E., Kennicutt R.~C., Goddard Q., 2009, ApJ, 695, 580
\bibitem[Bresolin et al.(2005)]{bresolin05} Bresolin F., Schaerer D., Gonz\'alez Delgado R. M., Stasi\'nska G., 2005, A\&A, 441, 981 
\bibitem[Caffau et al.(2016)]{caffau16} Caffau  E. et al. 2016, A\&A, 585, 16
\bibitem[Campbell(1988)]{campbell88} Campbell A., 1988, ApJ, 335, 644
\bibitem[Cacho et al.(2014)]{cacho14} Cacho R., S\'anchez-Bl\'azquez P., Gorgas J.,  P\'erez I., 2014, MNRAS, 442, 2496
\bibitem[Chopinet \& Lortet-Zuckermann(1976)]{chopinet76} Chopinet M., \& Lortet-Zuckermann M. C., 1976, A\&AS, 25, 179
\bibitem[Conti et al.(2007)]{conti07} 	Corti M.,  Bosch G., Niemela V., 2007, A\&A, 467, 137
\bibitem[Copetti et al.(2000)]{coppeti00} Copetti M.~V.~F., Casta\~neda H.~O., Mallmann J.~A.~H., Schmidt A.~A., 2000, A\&A, 357, 621
\bibitem[Dors et al.(2011)]{dors11}  Dors O.~L., Krabbe A.,  H\"agele G.~F., P\'erez-Montero  E., 2011, MNRAS, 415, 3616
\bibitem[Dors \& Copetti(2003)]{dors03} Dors O.~L., \& Copetti M.~V.~F., 2003, A\&A, 404, 969
\bibitem[Dors  \& Copetti(2005)]{dors05} Dors O.~L., \& Copetti M.~V.~F., 2005, A\&A, 437, 837
\bibitem[Dors  et al.(2008)]{dors08} Dors O.~L.,  Storchi-Bergmann T., Riffel R.~A., Schimdt  A.~A. 2008, A\&A, 482, 59
\bibitem[Dors  et al.(2013)]{dors13} Dors O.~L. et al., 2013, MNRAS, 432, 2512
\bibitem[D\'{\i}az et al.(2007)]{diaz07} D\'{\i}az A.~I., Terlevich E.,  Castellanos M.,  H\"agele  G.~F., 2007, MNRAS, 382, 251
\bibitem[D\'{\i}az et al.(2000)]{diaz00} D\'{\i}az A.~I., Castellanos M., Terlevich E., Gar\'{\i}cia-Vargas L.~M., 2000, MNRAS, 318, 462
\bibitem[D\'{\i}az et al.(1991)]{diaz91} D\'{\i}az A.~I., Terlevich E., V\'{\i}lchez J.~M., Pagel B.~E.~J., Edmunds M. G., 1991,
MNRAS, 253, 245
\bibitem[de Koter et al.(1997)]{dekoter97} de Koter A., Heap S.~R., Hubeny I., ApJ, 477, 792
\bibitem[Evans et al.(2015)]{evans15} Evans C. J., Kennedy M. B.,  Dufton P. L., 2015, A\&A, 574, 13
\bibitem[Evans(1986)]{evans86} Evans I.~N., 1986, ApJ, 309, 544
\bibitem[Evans \& Dopita(1985)]{evans85} Evans I.~N., \& Dopita M.~A., 1985, ApJS, 58, 125 
\bibitem[Esteban \& M\'endez(1999)]{esteban99} Esteban C., \& M\'endez D.~I., 1999, A\&A, 348, 446
\bibitem[Fierro et al.(1986)]{fierro86} Fierro J., Torres-Peimbert S., Peimbert M., PASP, 1986, 98, 1032
\bibitem[Ferland et al.(2013)]{ferland13} Ferland G. J. et al., 2013, Rev. Mex. Astron. Astrofis., 49, 137
\bibitem[Garnett et al.(1997)]{garnett97} Garnett D.~R., Shields G.~A., Skillman E.~D., Sagan S.~P., Dufour R.~J., 1997, ApJ, 489, 63
\bibitem[Garnett \& Kennicutt(1994)]{garnett94} Garnett D.~R., \& Kennicutt R.~C., 1994, ApJ, 426, 123
\bibitem[Gurzadyan(1955)]{gurzadyan55} Gurzadyan G. A. 1955, Soob. Burakan Obs., 16, 3
\bibitem[H{\"a}gele et al.(2013)]{hagele13} H{\"a}gele, G.~F., 
D{\'{\i}}az, {\'A}.~I., Terlevich, R., et al.\ 2013, MNRAS, 432, 810 
\bibitem[H\"agele et al.(2012)]{hagele12} H\"agele G.~F., Firpo V., Bosch G., D\'{\i}az  A.~I., Morrell N., 2012, MNRAS, 422, 3475
\bibitem[H\"agele et al.(2011)]{hagele11} H\"agele G.~F. et al., 2011, MNRAS, 414, 272
\bibitem[H{\"a}gele et al.(2010)]{hagele10} H{\"a}gele, G.~F., 
D{\'{\i}}az, {\'A}.~I., Cardaci, M.~V., Terlevich, E., 
\& Terlevich, R.\ 2010, MNRAS, 402, 1005 
\bibitem[H{\"a}gele et al.(2009)]{hagele09} H{\"a}gele, G.~F., 
D{\'{\i}}az, {\'A}.~I., Cardaci, M.~V., Terlevich, E., 
\& Terlevich, R.\ 2009, MNRAS, 396, 2295 
\bibitem[H{\"a}gele et al.(2008)]{hagele08} H{\"a}gele G.~F. et al.,  2008, MNRAS, 383, 209
\bibitem[H{\"a}gele et al.(2007)]{hagele07} H{\"a}gele, G.~F., 
D{\'{\i}}az, {\'A}.~I., Cardaci, M.~V., Terlevich, E., 
\& Terlevich, R.\ 2007, MNRAS, 378, 163 
\bibitem[H{\"a}gele et al.(2006)]{hagele06} H{\"a}gele G.~F., P{\'e}rez-Montero E., D{\'{\i}}az  A.~I., Terlevich E.,  Terlevich R.,  2006, MNRAS, 372, 293 
\bibitem[Henry \& Howard(1995)]{henry95} Henry R.~B.~C., \& Howard J.~W., 1995, ApJ, 438, 170
\bibitem[Hummer \& Storey(1997)]{hummer97} Hummer D. G., \& Storey P. J., 1987, MNRAS, 224, 8018
\bibitem[Iijima(1981)]{iijima81} Iijima T., 1981, in NATO Advanced Study Institute 69, Photom
etric and Spectroscopic Binary Systems, 517 
\bibitem[Kaler(1978)]{kaler78} Kaler J. B., 1978, ApJ, 220, 887
\bibitem[Kaler(1976)]{kaler76} Kaler J. B., 1976, ApJ, 210, 843
\bibitem[Koppen \& Tarafda(1978)]{koppen78} K\"oppen  J., \& Tarafdar S. P., 1978, A\&A, 69, 363
\bibitem[Kennicutt \& Garnett(1996)]{kennicutt96} Kennicutt D.~R., Garnett D.~R.,  1996, ApJ, 456, 504
\bibitem[Kennicutt et al.(2003)]{kenniccut03} Kennicutt R. C., Bresolin  F.,  Garnett D. R., 2003, ApJ, 591, 801
\bibitem[Kennicutt et al.(2000)]{kennicutt00} Kennicutt  R.~C.,  Bresolin F., French  H., Martin  P., 2000, ApJ, 537, 589
\bibitem[Kewley \& Ellison(2008)]{kewley08} Kewley L. J., \& Ellison S., 2008, ApJ, 681, 1183
\bibitem[Kewley \& Dopita(2002)]{kewley02} Kewley L.~J., \& Dopita M.~A., 2002, ApJS, 142, 35
\bibitem[Kewley et al.(2001)]{kewley01} Kewley L.~J., Dopita M.~A., Sutherland R.~S., Heisler C.~A., Trevena J., 2001, ApJ, 556, 121
\bibitem[Kwitter \& Aller(1981)]{kwitter81} Kwitter K.~B., \&  Aller L.~H., 1981, MNRAS, 195, 939
\bibitem[Krabbe et al.(2014)]{krabbe14} Krabbe A. C. et al., 2014, MNRAS, 437, 1155
\bibitem[Lamb et al.(2015)]{lamb15} Lamb J.~B. et al., 2015, arXiv:1512.01233v1
\bibitem[Lee \& Skillman(2004)]{lee04} Lee H., Skillman E. D., 2004, ApJ, 614, 698
\bibitem[Leitherer et al.(1999)]{leitherer99} Leitherer C. et al., 1999, ApJ, 123, 3 
\bibitem[L\'opez-S\'anchez et al.(2007)]{lopezsanchez07}  L\'opez-S\'anchez A.~R., Esteban C., Garc\'{\i}a-Rojas J., Peimbert M., Rodr\'{\i}guez M., 2007, ApJ, 656, 168
\bibitem[L\'opez-S\'anchez et al.(2011)]{lopezsanchez11} L\'opez-S\'anchez A.~R., Mesa-Delgado A., L\'opez-Martín L., Esteban C., 2011, MNRAS, 411, 2076
\bibitem[L\'opez-S\'anchez \& Esteban(2009)]{lopezsanchez09} L\'opez-S\'anchez A.~R., Esteban C.,  2009, A\&A, 508, 615
\bibitem[L\'opez-Hern\'andez et al.(2013)]{lopezhernadez13} L\'opez-Hern\'andez J. et al., 2013, MNRAS, 430, 472
\bibitem[Massey(2011)]{massey11} Massey  P., 2011,   in Astronomical Society of the Pacific
Conference Series, Vol. 440, UP2010: Have Observations
Revealed a Variable Upper End of the Initial Mass Function
ed. M. Treyer, T. Wyder, J. Neill, M. Seibert,\& J. Lee, 29
\bibitem[Massey et al.(2009)]{massey09} Massey  P., Zangari A.~M., Morrell N.~I., 2009, ApJ, 692, 618
\bibitem[Massey et al.(2005)]{massey05} Massey  P. et al. 2005, ApJ, 627, 477
\bibitem[Mayya \& Prabhu(1996)]{mayya96} Mayya  Y.~D., \&  Prabhu T.~P., 1996, AJ, 111, 1252
 \bibitem[Morisset(2004)]{morisset04} Morisset  C., 2004, ApJ, 601, 858
 \bibitem[Morisset et al.(2016)]{morisset16} Morisset  C., 2016,  arXiv160601146M
 \bibitem[Morrell et al.(2014)]{morrell14} Morrell N.~I.,  Massey P.,  Neugent K.~F. 2014, ApJ, 789, 139
\bibitem[Mathis(1985)]{mathis85} Mathis J. S., 1985, ApJ, 291, 247 
\bibitem[Oey et al.(2000)]{oey00} Oey M.~S., Dopita M.~A., Shields J.~C., Smith R.~C. 2000, ApJSS, 128, 511
\bibitem[Osterbrock(1989)]{osterbrock89} Osterbrock D. E., 1989, Astrophysics of Gaseous Nebulae and Active
Galactic Nuclei. University Science Books, Mill Valley, CA
\bibitem[Pagel et al.(1979)]{pagel79} Pagel B.~E.~J., Edmunds M.~G., Blackwell D.~E., Chun M.~S.,  Smith G., 1979, MNRAS, 189, 95
\bibitem[Pauldrach et al.(2001)]{pauldrach01} Pauldrach A.~W.~A., Hoffmann T.~L.,  Lennon M., 2001, A\&A, 375, 161
\bibitem[Pellegrini, Baldwin, \& Ferland(2011)]{pellegrini11} Pellegrini E.~W., Baldwin J.~A., Ferland G.~J., 2011,  ApJ, 738, 34
\bibitem[Pellegrini, Baldwin, \& Ferland(2010)]{pellegrini10} Pellegrini E.~W., Baldwin J.~A., Ferland G.~J., 2010, ApJSS, 191, 160
\bibitem[Pe\~na-Guerrero et al.(2012)]{pena12} Pe\~na-Guerrero M.~A., Peimbert A., Peimbert M., 2012, ApJ, 756, 14
\bibitem[P\'erez-Montero(2014)]{perez14} P\'erez-Montero  E., 2014, MNRAS, 441, 2663
\bibitem[P\'erez-Montero \& V\'{\i}lchez(2009)]{enrique09a} P\'erez-Montero  E., \&  V\'{\i}lchez  J.~M., 2009, MNRAS, 400, 1721
\bibitem[P\'erez-Montero \& Contini(2009)]{perez09} P\'erez-Montero  E., \& Contini T., 2009, MNRAS, 398, 949
\bibitem[P\'erez-Montero \& D\'{\i}az(2006)]{enrique06} P\'erez-Montero  E., \& D\'{\i}az A.~I. 2006, MNRAS, 449, 193
\bibitem[Pettini \& Pagel(2004)]{pettini04} Pettini M.,  \& Pagel B.~E.~J., 2004, MNRAS, 348, 59
\bibitem[Pilyugin et al.(2012)]{leonid12} Pilyugin  L. S., Grebel E. K., Mattsson L., 2012, MNRAS, 424, 2316
\bibitem[Pilyugin \& Grebel(2016)]{pilyugin16} Pilyugin L. S., \&  Grebel E. K., 2016, MNRAS, 457, 3678
\bibitem[Pilyugin(2001)]{leonid01} Pilyugin  L. S.,  2001, A\&A, 369, 594
\bibitem[Pilyugin(2000)]{leonid00} Pilyugin  L. S.,  2000, A\&A,  362, 325
\bibitem[Preite-Martinez \& Pottasch(1983)] Preite-Martinez  A., \& Pottasch S. R., 1983, A\&A, 126, 31
\bibitem[Rosa et al.(2014)]{rosa14} Rosa D.~A. et al., 2014, MNRAS, 444, 2005
\bibitem[Russel \& Dopita(1990)]{russel90} Russel S.~S., \& Dopita M.~A., 1990, ApJS, 74, 93
\bibitem[Sanders et al.(2016)]{sanders16} Sanders R.~L. et al., 2016, ApJ, 816, 23
\bibitem[S\'anchez et al.(2012)]{sanchez12} S\'anchez S. F. et al., 2012, A\&A, 546, 2 
\bibitem[Schaerer \& de Koter(1997)]{schaerer97}  Schaerer D., \& de Koter A., 1997, A\&A, 322, 598 
\bibitem[Shields \& Searle(1976)]{shields76} Shields G. A., \& Searle L., 1978, ApJ, 222, 821 
\bibitem[Skillman \& Kennicutt(1993)]{skillman93} Skillman E.~D., \& Kennicutt R.~C., 1993, ApJ, 411, 655
\bibitem[Salpeter(1955)]{salpeter55} Salpeter E.~E., 1955, ApJ, 121, 161
\bibitem[Sota et al.(2014)]{sota14} Sota  A.,  Ma\'{\i}z  Apell\'aniz  J.,  Morrell N.~I. et al., 2014, ApJS, 211, 10
\bibitem[Sota et al.(2011)]{sota11} Sota  A., Ma\'{\i}z  Apell\'aniz  J., Walborn N. R., 2011, ApJS, 193, 24
\bibitem[Stasi\'nska \& Tylenda(1986)]{stasinska86} Stasinska G., \& Tylenda R., 1986, A\&A, 155, 137
\bibitem[Stasi\'nska(1982)]{stasinska82} Stasinska G., 1982, A\&ASS, 84, 320
\bibitem[Storey \& Zeippen(2000)]{storey00} Storey P.~J., \& Zeipper C.~J., 2000, MNRAS, 312, 813
\bibitem[Stoy(1933)]{stoy33} Stoy R. H., 1933, MNRAS, 93, 588 
\bibitem[Vacca et al.(1996)]{vacca96} Vacca W. D., Garmany C. D.,  Shull J. M., 1996, ApJ, 460, 914
\bibitem[van Hoof et al.(2001)]{vanhoof01} van Hoof, P. A. M., Weingartner, J. C., Martin, P. G., Volk, K., \&
Ferland, G. J. 2001, in Challenges of Photoionized Plasmas, ed. G. Ferland \& D. Savin (San Francisco: ASP), ASP Conf Ser., 247, 363
\bibitem[Van Zee et al.(2000)]{vanzee00} van Zee  L., 2000, ApJ, 543, L31
\bibitem[Van Zee et al.(1998)]{vanzee98}van Zee  L., Salzer J.~J., Haynes M.~P., O'Donoghue A.~A., Balonek T.~J., 1998, AJ, 116, 2805
\bibitem[Veilleux \& Osterbrock(1987)]{veilleux87} Veilleux S., Osterbrock D.~E. 1987, ApJS, 63, 295
\bibitem[Vermeij et al.(2002)]{vermeij02} Vermeij R., Damour F., van der Hulst J.~M., Baluteau J.-.P, 2002, A\&A, 390, 649
\bibitem[Vila Costas \& Edmunds(1993)]{vilacostas93}	Vila-Costas, M. B., \& Edmunds, M. G. 1993, MNRAS, 265, 199 
\bibitem[V\'{\i}lchez \& Pagel(1988)]{vilchez88} V\'{\i}lchez J.~M., \&  Pagel B.~E.~J., 1988, MNRAS, 231, 257 
\bibitem[V\'{\i}lchez et al.(1988)]{vilchez88a} V\'{\i}lchez J.~M., Pagel B.~E.~J., D\'{\i}az  A.~I., Terlevich E.,  Edmunds M. G., 1988, MNRAS, 235, 633
\bibitem[Zanstra(1931)]{zanstra31} Zanstra, H. 1931, Publ. Dominion Astrophys. Obs., 4, 209
\bibitem[Zastrow et al.(2013)]{zanstrow13} Zastrow, J.,  Oey M.~S.,  Pellegrini E.~W., 2013, ApJ, 769, 94
\bibitem[Zurita \& Bresolin(2012)]{zurita12} Zurita A., \&  Bresolin F., 2012, MNRAS, 427, 1463		
\bibitem[Walborn et al.(2014)]{walborn14} Walborn N. R., Sana H., Sim\'on-D\'{\i}az S., 2014, A\&A, 564, 40
\end{thebibliography}
\end{document}